\documentclass[aps,prd,nofootinbib]{revtex4-1}
\usepackage{graphicx}				
\usepackage{epsfig}
\usepackage{bm}
\usepackage{amssymb}
\usepackage{amsmath}
\usepackage{color}			
\usepackage{amssymb}
\newcommand{\be}{\begin{equation}}
\newcommand{\ee}{\end{equation}}
\newcommand{\bea}{\begin{eqnarray}}
\newcommand{\eea}{\end{eqnarray}}

\newcommand{\eq}[1]{Eq.~(\ref{#1})}

\newcommand{\dd}{{\rm d}}
\newcommand{\kp}{{{\rm k}_\perp}}
\newcommand{\vKT}{\vec K_\perp}

\newcommand{\bcl}[3]{b^\dagger_{#1,#2,#3}}

\newcommand{\bal}[3]{b_{#1,#2,#3}}
\newcommand{\delc}[4]{\delta_{#1,#2}^{#3 #4}\,}
\newcommand{\delcl}[6]{\delta_{#1,#2}^{#3 #4,#5 #6}\,}
\newcommand{\del}[2]{\delta_{#1,#2}\,}
\newcommand{\id}{1\!\!1}
\newcommand{\tr}{\mathrm{tr\, }}

\newcommand{\sumint}[2]{\int\limits_{#1}^{#2}\!\!\!\!\!\!\!\!\!\sum}
\newcommand{\bra}[1]{\Bigl<#1\Bigl|}
\newcommand{\ket}[1]{\Bigr|#1\Bigr>}
\newcommand{\trho}{{\tilde \rho}}
\def\cG{{\cal G}}
\def\la{\langle}\def\ra{\rangle}
\def\xp{x_\perp}\def\yp{y_\perp}
\def\kp{k_\perp}

\def\lsim{\mathrel{\rlap{\lower4pt\hbox{\hskip1pt$\sim$}}
    \raise1pt\hbox{$<$}}}                
\def\gsim{\mathrel{\rlap{\lower4pt\hbox{\hskip1pt$\sim$}}
    \raise1pt\hbox{$>$}}}                

\def\simge{\mathrel{%
   \rlap{\raise 0.511ex \hbox{$>$}}{\lower 0.511ex \hbox{$\sim$}}}}
\def\simle{\mathrel{
   \rlap{\raise 0.511ex \hbox{$<$}}{\lower 0.511ex \hbox{$\sim$}}}}
\def\bigs{\mathrel{
   \rlap{\raise 0.531ex \hbox{$>$}}{\lower 0.531ex \hbox{$<$}}}}

\newcommand{\xx}{{\rm x}}

\begin{document}
\hskip13cm {\bf NT@UW-18-08}
\title{Extracting many-body color charge correlators in the proton\\ from exclusive DIS at large Bjorken x}

\author{Adrian Dumitru}
\email{adrian.dumitru@baruch.cuny.edu}
\affiliation{Department of Natural Sciences, Baruch College,
CUNY, 17 Lexington Avenue, New York, NY 10010, USA}
\affiliation{The Graduate School and University Center, The City
  University of New York, 365 Fifth Avenue, New York, NY 10016, USA}
\affiliation{Physics Department, Brookhaven National Laboratory, Bldg. 510A, Upton, NY 11973, USA}

\author{Gerald A. Miller}
\email{miller@uw.edu}
\affiliation{Department of Physics, University of Washington, Seattle, WA 98195-1560, USA}

\author{Raju Venugopalan}
\email{raju@bnl.gov}
\affiliation{Physics Department, Brookhaven National Laboratory, Bldg. 510A, Upton, NY 11973, USA}

\date{\today}

\begin{abstract}
We construct a general QCD light front formalism to compute many-body
color charge correlators in the proton. These form factors can be
extracted from deeply inelastic scattering measurements of exclusive
final states in analogy to electromagnetic form factors extracted in
elastic electron scattering experiments. Particularly noteworthy is
the potential to extract a novel Odderon form factor, either
indirectly from exclusive $J/\Psi$ measurements, or directly from
exclusive measurements of the $\eta_c$ or tensor mesons at large
Bjorken x. Besides the intrinsic information conveyed by these color
charge correlators on the spatio-temporal tomography at the
sub-femtoscopic scale at large x, the corresponding cumulants extend
the domain of validity of McLerran-Venugopalan type weight functionals
from small x and large nuclei to nucleons and light nuclei at large
$x$, as well as to non-zero momentum transfer. This may
significantly reduce nonperturbative systematic uncertainties in the
initial conditions for QCD evolution equations at small $x$ and could
be of strong relevance for the phenomenology of present and future
collider experiments.

\end{abstract}

\maketitle

\section{Introduction} 

The increasing availability of high energies and high luminosities at fixed target and collider experiments~\cite{Boer:2011fh,Dudek:2012vr}
 allows for unprecedented access to the internal transverse spatial and momentum distributions of  color charge distributions inside nucleons and in nuclei.
 The standard framework~\cite{Ji:2003ak} is that of Wigner distributions~\cite{Wigner:1932eb} that allow simultaneous knowledge of both  spatial and momentum aspects of the nucleon wave function. Knowledge of the Wigner distributions allows the construction of 
   generalized parton distributions (GPDs)~\cite{Mueller:1998fv,Ji:1996ek,Radyushkin:1996nd,Collins:1996fb,Diehl:2003ny,Belitsky:2005qn,Burkardt:2002hr} and  transverse momentum distributions (TMDs)~\cite{Ralston:1979ys,Collins:1981uk,Mulders:1995dh,Boer:1997nt,Belitsky:2002sm,Miller:2008sq} that are generalizations of the usual collinear parton distributions. The GPDs provide information on the spatial tomography of the nucleon and TMDs allow for its momentum tomography.\\
   
These various distributions are very valuable. Our aim here is to
introduce a complementary approach employing the Hamiltonian light
front formalism in light cone gauge that allows essential insight into
the dynamics of color charges in nucleons and nuclei.  In this
framework, color charge densities, and higher cumulants of these, can
be defined and expressed as matrix elements of nonperturbative
boost-invariant light front Fock-space wave functions of the QCD
Hamiltonian. The corresponding form factors can be related to physical
observables; these are the exclusive final states measured in deeply
inelastic scattering (DIS) experiments. The information on color
charge distributions extracted from such exclusive DIS measurements
will be closely analogous to the information gathered on electric
charge and magnetization distributions from form factors measured in
elastic scattering of electrons by nucleons and
nuclei~\cite{Punjabi:2015bba,Miller:2007uy,Miller:2010nz,Carlson:2007xd}.\\

However because the QCD coupling $\alpha_S$ is much stronger than the QED fine structure constant, exclusive DIS experiments provide more
information on color charge distributions, and higher cumulants of these, than elastic scattering experiments. Though it is true that GPDs and
TMDs can be expressed in terms of light front wave functions~\cite{Burkardt:2000za,Dominguez:2011wm,Petreska:2018cbf}, our treatment in terms of 
 color charge densities is novel.\\
 
The suite of feasible exclusive DIS final states is a rich source of
information on many-body parton correlations with variations in
$x_{\rm Bj}$ and $Q^2$ and can be expected to lead to an understanding
of the internal spatial color charge structure of nucleons.  The
possible modification of this structure in nuclei, could be important
for understanding the EMC effect in DIS and nucleon-nucleon short
range correlations in nuclei~\cite{Hen:2016kwk,Miller:2015tjf}. Also very intriguing
is the possibility of comparing the color charge form factors to be
discussed here with those that are now beginning to be extracted from
lattice QCD computations~\cite{Winter:2017bfs}.\\

An attractive feature of the Hamiltonian light front framework is that
the color charge form factors extracted in DIS can be employed to
compute cross sections in hadron-hadron and hadron-nucleus
scattering. The usefulness of such color charge form factors is known
for QCD in the Regge limit of high energy scattering, with momentum
resolution scales $Q^2 \gg \Lambda_{\rm QCD}^2$ and $x_{\rm Bj} \sim
Q^2/s\rightarrow 0$, with $s$ representing the squared center of mass
energy in the experiment, as understood in the Color Glass Condensate
(CGC)~\cite{Iancu:2003xm,Gelis:2010nm,Kovchegov:2012mbw,Blaizot:2016qgz}. This
is an effective field theory of the Regge limit of QCD that is
formulated on the light front, with all the nontrivial information
regarding multigluon correlations contained in a gauge invariant
weight functional $W[\rho]$ that plays the role of a density
matrix. Here, $\rho$ represents the color charge density of large $x$
partons coupled to small $x$ gluon fields.\\

This weight functional was first derived by McLerran and Venugopalan (MV)~\cite{McLerran:1993ni,McLerran:1993ka,McLerran:1994vd}, who also outlined the elements of the CGC EFT using light front arguments. They argued that for a large nucleus $A$, a probe of transverse size $\sim 1/Q$ couples coherently (for $x\ll A^{-1/3}$) along its path length to partons confined to nucleons in the nucleus. While on average, the probe sees no net color charge, the physics of random walks indicates that it will see large fluctuations of the color charge and therefore, by the central limit theorem,  $W[\rho]$ will be Gaussian. These statements can be formulated 
with mathematical rigor~\cite{Kovchegov:1996ty,Jeon:2004rk}.\\

The variance of the Gaussian, is the color charge squared per unit
area $\mu_{\rm MV}^2\propto A^{1/3}$.  In the large $A$ limit
$\mu_{\rm MV}^2 \gg \Lambda_{\rm QCD}^2$, so that the CGC is a weakly
coupled EFT that allows for systematic computation of multigluon
correlation functions that capture the physics underlying the
phenomenon of gluon saturation~\cite{Gribov:1984tu,Mueller:1985wy} in
the high energy limit.  The building block of gluon radiation, the
Weizs\"{a}cker-Williams distribution, is screened at the scale
$Q_S^2\propto \mu_{\rm
  MV}^2$~\cite{Ayala:1995kg,JalilianMarian:1996xn,Kovchegov:1996ty},
and one recovers the phenomenologically successful Glauber-Mueller
dipole model~\cite{Mueller:1989st,Mueller:1999wm} of gluon 
saturation~\cite{McLerran:1998nk,Kovchegov:1999yj,Venugopalan:1999wu}.\\

The MV model does not describe the small $x$ evolution of the color source densities that arise from the $\alpha_S\ln(x) \sim O(1)$ enhanced bremsstrahlung of gluons.  This is given by
 the JIMWLK equation that describes the functional renormalization group evolution of $W[\rho]$ with decreasing $x$ 
 ~\cite{JalilianMarian:1997gr,JalilianMarian:1997dw,Iancu:2000hn,Ferreiro:2001qy}. This functional equation gives the Balitsky-JIMWLK hierarchy~\cite{Balitsky:1995ub,Weigert:2000gi}. The equivalent  functional Langevin equation  was solved numerically~\cite{Blaizot:2002np,Rummukainen:2003ns}. In the limit of large $N_c$, and large $A$, the lowest equation in this hierarchy, describing the $x$ evolution of ``dipole" 2-point correlators of lightlike Wilson lines, has a closed form expression--the Balitsky-Kovchegov (BK) equation~\cite{Balitsky:1995ub,Kovchegov:1999yj}, which reduces to  the BFKL equation~\cite{Kuraev:1977fs,Balitsky:1978ic} if  the density of sources is sufficiently low.\\
 
Remarkably, as first conjectured in \cite{Iancu:2002aq}, numerical simulations of the functional Langevin equation demonstrate that the hierarchy of correlators is to good approximation solved by a Gaussian $W[\rho]$~\cite{Dumitru:2011vk}, with  $\mu_{\rm MV}^2\longrightarrow \mu_{\rm JIMWLK}^2(x,\kp)$, where $ \mu_{\rm JIMWLK}^2(x,\kp)$ is given by the solution of the BK equation. This Gaussian effective theory provides a quantitative phenomenology of electron-proton collisions at HERA~\cite{Albacete:2010sy,Kuokkanen:2011je,Lappi:2013zma}. Further, the formulation of the CGC EFT in the language of color source densities allows a first principles formulation of multiparticle production in QCD at small $x$~\cite{Gelis:2006yv,Gelis:2006cr,Gelis:2007kn,Gelis:2008rw,Gelis:2008ad,Kovchegov:1998bi,Dumitru:2001ux}. \\

The initial conditions for BK/JIMWLK evolution are given by the MV model which, as noted, is formulated for large nuclei.  Here we are concerned with the nucleon at large $x$. In this
case, the central limit theorem is not applicable and the color charge form factors of the proton
can reasonably be expected to be very different than in the MV model.  Therefore a first principles computation of these form factors is in order. Such a
computation is of intrinsic interest and can help constrain the systematic uncertainties in the QCD evolution of color charge
distributions in the proton arising from the initial conditions. The spatial distributions of color charge density in the proton is also of
great topical interest because of the unexpected long range azimuthally collimated ``ridge" multiparticle correlations measured at
RHIC and LHC~\cite{Dusling:2015gta}; the latter may depend sensitively on the former~\cite{Bjorken:2013boa,Schenke:2014zha,McLerran:2016snu,Dusling:2018hsg,Boer:2018vdi,Kovchegov:2018jun,Mace:2018vwq}. Several
models have been constructed to incorporate spatial nucleon color charge distributions in describing these data. However, they are
constrained in varying degrees by systematic uncertainties in the initial conditions~\cite{Mantysaari:2016ykx,Bozek:2016kpf,Albacete:2016pmp}.\\

Here we  develop a light front Hamiltonian framework that can be used to compute color charge form factors in nucleons and nuclei. The light front formalism we will employ is standard; see  for instance  \cite{Brodsky:1997de}. We  focus on the simple problem of constructing quadratic and cubic combinants of a three quark Fock state at large $x$. The color charge combinants can alternatively be expressed in terms of color charge form factors. We will discuss how information on these form factors can be cleanly extracted in exclusive DIS measurements of vector and tensor mesons at large $x$. An interesting possibility is the extraction of a novel Odderon color charge form factor in such measurements~\cite{Lukaszuk:1973nt}. As we will discuss, large $x$ DIS exclusive measurements should be particularly sensitive to the Odderon. This is of topical interest in light of recent claims that the TOTEM experiment may have found evidence of Odderon exchange in proton-proton elastic scattering at the highest LHC energies~\cite{Antchev:2017yns}. \\

This paper is organized as follows. In section 2, we begin by
displaying the light front wavefunction for the proton, focusing
immediately on the three valence quark component of the wavefunction.
The extension to higher Fock states would be straightforward, but more
involved. We also establish the notations and conventions to be
employed in the rest of the paper. We then develop in section 3, in
successive subsections, the general framework to compute light front
color charge densities for the valence states, and the computation of
the expectation values of quadratic and cubic color charge
operators. In the last of these subsections, we compare our results to
the MV model and demonstrate the relation between the gluon
distribution in the proton and a quadratic correlator of color charge
densities. The relation of the corresponding color charge form factors
to exclusive heavy quark pair production in DIS is discussed in
Section 4. In particular, we show that $J/\Psi$ production is
sensitive to both a quadratic ``Pomeron" color charge form factor and
the cubic Odderon color charge form factor. In contrast, $\eta_c$ or
tensor meson production are depends only on the Odderon form factor.
In the concluding section, we will further discuss the prospects of
Odderon discovery in DIS experiments in light of prior searches. We
will also discuss more generally the prospects for quantitative
constraints on the quadratic and cubic color charge form factors from
DIS data at large $x_{\rm Bj}$. We shall also outline the next steps
both on further theoretical development of this framework and in
quantitative comparison and predictions for measurements at extant and
future experiments. The paper contains two appendices.  In Appendix A,
we discuss the color charge density operator in the limit of large
longitudinal momenta. In Appendix B, we provide some details of the
computation of the Odderon form factor.

\section {The light front proton wavefunction: notation and conventions}

In this section, we shall introduce our notation and conventions for
the proton wavefunction on the light front. These closely follow
Refs.~\cite{Brodsky:1989pv,Brodsky:2000ii}. The light front
wavefunction of an unpolarized on-shell proton with four-momentum
$P^\mu = (P^+, P^-,\vec{P}_\perp)$ can be expressed as
\be
|P\rangle = \int \dd \mathrm{PS}_n \sum_n \psi_n |n\rangle~\,,
\ee
where $|n\rangle$ are the Fock space basis vectors of the light front
Hamiltonian, $\psi_n = \langle n|P\rangle$ is the amplitude for a particular
Fock state $|n\rangle$ in the proton and $\dd \mathrm{PS}$ denotes the $n$-body phase space for $|n\rangle$. 
If the proton light front wavefunction is dominated by its valence quark state, as is the case at large values of Bjorken $x$, it is given
explicitly as 
\bea
|P\rangle &=& \frac{1}{\sqrt{6}} \int \frac{\dd x_1\dd x_2 \dd x_3}{\sqrt{x_1 x_2 x_3}}
\delta(1-x_1-x_2-x_3)
\int \frac{\dd^2 k_1 \dd^2 k_2 \dd^2 k_3}{(16\pi^3)^3}
 16\pi^3 \delta(\vec{k}_1+\vec{k}_2+\vec{k}_3)\nonumber\\
  &\times& 
 \sum_{\lambda_1,\lambda_2,\lambda_3}
 \psi_3(p_1, \lambda_1,p_2, \lambda_2, p_3, \lambda_3)
 \sum_{i_1, i_2, i_3}\epsilon_{i_1 i_2 i_3}
  |p_1,i_1, \lambda_1; \, p_2,i_2, \lambda_2; \, p_3,i_3, \lambda_3\rangle~.  \label{eq:def_|P>}
\label{eq:valence-proton}
\eea
The three on-shell quark momenta are specified by their lightcone
momenta $p_i^+ = x_i P^+$ and their transverse momenta\footnote{For a
  lighter notation we often suppress the $\perp$ subscript on quark
  transverse momenta.}  $\vec{p}_{i} = x_i \vec{P}_\perp +
\vec{k}_i$. Hence the $\vec{k}_i$ can be interpreted as the transverse
momenta of the valence quarks relative to the proton. In addition to
color, denoted by $i_i$, the quark Fock state also carries flavor and
helicity quantum numbers which are collectively denoted as
$\lambda_i$. The valence Fock state wave function in color space
belongs to the product space obtained from the direct product of three
triplet color spaces: $|i_1; i_2; i_3\rangle = |i_1\rangle \otimes
|i_2\rangle \otimes |i_3\rangle$.  The Levi-Civita tensor in
\eq{eq:def_|P>} projects the product of three fundamental
representations onto the totally anti-symmetric SU(3) singlet; a SU(3)
transformation $U$ of $\epsilon_{i_1 i_2 i_3} |i_1\rangle
|i_2\rangle i_3\rangle$ gives
\be
\epsilon_{i_1 i_2 i_3} U_{j_1 i_1}U_{j_2 i_2}U_{j_3
  i_3}|j_1\rangle|j_2\rangle |j_3\rangle =
\epsilon_{j_1 j_2 j_3} (\mathrm{det~} U) |j_1\rangle|j_2\rangle |j_3\rangle ~,
\ee
where det~$U=1$ for $U\in$~SU(3).\\

The amplitude $\psi_3$ in Eq.~(\ref{eq:valence-proton}) is symmetric under exchange of any two of its
arguments and is normalized according to
\be \label{eq:Norm_psi3}
 \int {\dd x_1\dd x_2 \dd x_3}\, \delta(1-x_1-x_2-x_3)
  \int \frac{{\dd^2 k_1 \dd^2 k_2 \dd^2 k_3}}{(16\pi^3)^3}
  (16\pi^3)\,\delta(\vec{k}_1+\vec{k}_2+\vec{k}_3)\, 
 \sum_{\lambda_1,\lambda_2,\lambda_3}
  |\psi_3|^2 = 1~.
\ee
Note that $\psi_3$ vanishes when the set $\{\lambda_1, \lambda_2,
\lambda_3 \}$ does not match the corresponding quantum numbers of the
proton.  The normalization of $\psi_3$ corresponds to the proton
wavefunction normalization,
\bea
\langle K | P\rangle &=& 16\pi^3 \, P^+ \delta(P^+ - K^+)
\, \delta(\vec{P}_\perp - \vec{K}_\perp) \label{eq:ProtonNorm1} \\
&=& 16\pi^3 \, \delta(\Delta x)
\, \delta(\vec{P}_\perp - \vec{K}_\perp)~.  \label{eq:ProtonNorm2}
\eea
For simplicity, throughout the manuscript we take the fractional plus
momentum transfer $\Delta x = (K^+ - P^+)/P^+$ to be very small or zero.\\

The one-particle quark states introduced above are created by the action of the quark creation  operator $\bcl{p}{i}{\lambda}$ on the
one-particle vacuum $|0\rangle$:
\be
|p,i,\lambda\rangle =   \bcl{p}{i}{\lambda} |0\rangle~.
\ee
Its Hermitian conjugate transforms an occupied one-particle state to the light front vacuum state,
\bea
\bal{k}{j}{\sigma} |p,i,\lambda\rangle &=&   \delta^{ji}\delta^{\sigma\lambda}
\, k^+ \delta(k^+ - p^+)\, 16\pi^3
\delta(\vec{k} - \vec{p}) |0\rangle \equiv \delcl{k}{p}{j}{i}{\sigma}{\lambda} |0\rangle~,
   \label{eq:def_annihil}\\
\bal{k}{j}{\sigma} |0\rangle &=& 0~.
\eea
In \eq{eq:def_annihil}, we  introduced a short hand notation 
$\delcl{k}{p}{j}{i}{\sigma}{\lambda}$, which we will frequently use
throughout the rest of the paper. We shall further also use the shorthand notation, 
\bea
\delc{k}{p}{i}{j} &\equiv& \delta^{ji}\, \del{k}{p}\, \\
\del{k}{p} &\equiv& k^+
   \delta(k^+ - p^+)\, 16\pi^3 \delta(\vec{k} - \vec{p})~.
\eea\\

The quark creation and destruction  operators satisfy the anti-commutation relation,
\be
\{ \bal{k}{j}{\sigma} , \bcl{p}{i}{\lambda} \} = \delcl{k}{p}{j}{i}{\sigma}{\lambda}~.\label{anticomm}
\ee
These relations, along with the convention that $\langle
0|0\rangle=1$, determine the normalization of one-particle states as
\be
\langle k,j,\sigma | p,i,\lambda\rangle = \langle 0| \bal{k}{j}{\sigma}
\, \bcl{p}{i}{\lambda} |0\rangle
= \langle 0| \{ \bal{k}{j}{\sigma} , \bcl{p}{i}{\lambda} \} |0\rangle
=\delcl{k}{p}{j}{i}{\sigma}{\lambda}~.
\ee
Furthermore,
\be \label{eq:bc_ba_quark}
\langle k,j,\sigma | \bcl{q}{m}{\sigma'} \bal{r}{n}{\lambda'} |p,i,\lambda\rangle
= \delcl{k}{q}{j}{m}{\sigma}{\sigma'}\, \delcl{r}{p}{n}{i}{\lambda}{\lambda'}~,
\ee
and
\be \label{eq:ba_bc_quark}
\langle k,j,\sigma | \bal{q}{m}{\sigma'} \bcl{r}{n}{\lambda'} |p,i,\lambda\rangle
= \delcl{q}{r}{m}{n}{\sigma'}{\lambda'} \delcl{k}{p}{j}{i}{\sigma}{\lambda}
- \delcl{k}{r}{j}{n}{\sigma}{\lambda'}\, \delcl{q}{p}{m}{i}{\sigma'}{\lambda}~.
\ee
With these relations in hand, one can  derive  matrix elements of density operators and powers thereof.\\

Before we discuss color charge densities, let us first consider the following operator: 
\be
\left[\trho^{mn}_q\right]_1 = \sumint{\ell,\lambda}{} \, \bcl{\ell-q}{m}{\lambda}
\bal{\ell}{n}{\lambda} 
~.
\label{eq:number-operator}
\ee
We have written the integration measure here compactly as
\be \label{eq:Def_sumint}
\sumint{\ell,\lambda}{}  \equiv
\int\limits_0^\infty \frac{\dd\ell^+}{\ell^+} \int\frac{\dd^2
  \ell}{16\pi^3}\sum_\lambda ~~~~,~~~~
\sumint{\ell,\lambda}{} \delcl{q}{l}{i}{j}{\sigma}{\lambda} = \delta^{ij} ~.
\ee
Setting $\vec{P}_\perp=0$ in the incoming proton for simplicity,
and employing \eq{eq:bc_ba_quark} and \eq{anticomm}, we obtain the
expectation value of the operator defined in \eq{eq:number-operator} as 
\bea
\langle K|\left[\trho^{mn}_q\right]_1 | P \rangle &=& \frac{1}{16\pi^3} \delta^{mn}
 \int \frac{\dd x_1\dd x_2 \dd x_3}{\sqrt{x_1 x_2 x_3}}
\delta(1-x_1-x_2-x_3)
\int {\dd^2 p_1 \dd^2 p_2 \dd^2 p_3}
\, \delta(\vec{p}_1+\vec{p}_2+\vec{p}_3)\nonumber\\
&\times & 
 \int \frac{\dd y_1\dd y_2 \dd y_3}{\sqrt{y_1 y_2 y_3}}
\delta(1-y_1-y_2-y_3)
\int {\dd^2 k_1 \dd^2 k_2 \dd^2 k_3}
\, \delta(\vec{k}_1+\vec{k}_2+\vec{k}_3) \nonumber\\
&\times & 
\sum_{\lambda_1,\lambda_2,\lambda_3}
\psi_3^*(k_1,k_2,k_3)\, \psi_3(p_1, p_2, p_3)
\, \del{k_1}{p_1-q} \del{k_2}{p_2} \del{k_3}{p_3}~.
\eea
It is implied that $y_i$, $\vec{k}_i$ are the momentum fractions and
transverse momenta, respectively, of the quarks in the outgoing
proton.  However, there is a subtlety: the plus momenta of the quarks
in the outgoing proton correspond to $k_i^+ = y_i K^+ = y_i (1+\Delta
x) P^+$ rather than to $k_i^+ = y_i P^+$. Therefore, in the arguments
of the delta-functions originating from the Fock space matrix elements
(the last three in the expression above) we have to shift $y_i \to y_i
(1+\Delta x)$; we also have to shift $\vec{k}_i \to \vec{k}_i + y_i
\vec{K}_\perp$ since there is a non-zero transfer of transverse
momentum.  To simplify the final expression we shall take $\Delta
x\to0$ so that
\bea
\langle K|\left[ \trho^{mn}_q\right]_1 | P \rangle &=& {16\pi^3}
\delta(\vec{q}+\vec{K}_\perp) \,
\delta(x_q+\Delta x) \delta^{mn}\,\nonumber\\
&\times &  \int \dd x_1\dd x_2 \dd x_3 \, \delta(1-x_1-x_2-x_3) 
\int {\dd^2 p_1 \dd^2 p_2 \dd^2 p_3\over (16\pi^3)^2}
\, \delta(\vec{p}_1+\vec{p}_2+\vec{p}_3) \nonumber\\
&\times & \sum_{\lambda_1,\lambda_2,\lambda_3}
\psi_3^*(k_1,k_2,k_3)\, \psi_3(p_1, p_2, p_3)~.
\label{eq:rho_mn}
\eea
In the limit $\Delta x\to0$ the arguments of $\psi_3^*$ are
$k_i^+\simeq x_iP^+$, $\vec{k}_1 \simeq \vec{p}_1+(1-x_1)\vec{K}_\perp$,
$\vec{k}_2 \simeq \vec{p}_2-x_2\vec{K}_\perp$,
$\vec{k}_3 \simeq \vec{p}_3-x_3\vec{K}_\perp$. (Note that the flavor and helicity
of each quark remains unchanged.)\\

The prefactor, $ {16\pi^3} \delta(\vec{q}+\vec{K}_\perp) \,
\delta(x_q+\Delta x)$,  of \eq{eq:rho_mn} is the overlap $\langle K|P\rangle$.  This factor enters in the matrix elements that we compute, but according to the usual Feynman rules do not appear in the final invariant amplitudes.  The remaining factors are $\delta^{mn}$ and a  dimensionless matter ($M$) form factor, $F_M(q)$:
\bea F_M(q)\equiv  \int \dd x_1\dd x_2 \dd x_3 \, \delta(1-x_1-x_2-x_3) 
\int {\dd^2 p_1 \dd^2 p_2 \dd^2 p_3\over (16\pi^3)^2}
\, \delta(\vec{p}_1+\vec{p}_2+\vec{p}_3) 
 \sum_{\lambda_i}
\psi_3^*(k_1,k_2,k_3)\, \psi_3(p_1, p_2, p_3)~.
\label{eq:rho_mn1}
\eea

If the transverse momentum transfer $\vec{K}_\perp$ is also much
smaller than the typical momenta of the quarks in the proton, the
remaining integral is proportional to the normalization integral for
$\psi_3$ given in \eq{eq:Norm_psi3}. In that case,
\bea
\langle K|\left[ \trho^{mn}_q \right]_1| P \rangle \simeq 16\pi^3
\, \delta^{mn}\, \delta(\vec{q}+\vec{K}_\perp)
\, \delta(x_q+\Delta x)~.
\eea
Indeed, stripping off the color space identity matrix and setting both
$x_q$ and $\vec{q}$ to zero leads back to the normalization
condition in \eq{eq:ProtonNorm2} for the proton wavefunction.

\section{Light front expectation values of color charge densities and form factors}

After the prior discussion of the essential preliminaries, we have all the elements in place to construct the light front color charge operator and expectation values of 
moments of expectation values of this operator in the large $x$  kinematic region where valence quarks dominate. We will later discuss the relation of these correlators to 
cross-sections for exclusive DIS final states. 

\subsection {The color charge density operator}

The color charge current density associated
with $f=1\dots N_f$ fermion fields $\psi_f$ is $j^{\mu a} =
\bar\psi_{i,f} \gamma^\mu \psi_{j,f} (t^a)_{ij}$. Here, $t^a$, $a=1\dots8$
are the generators of the fundamental representation of color-SU(3)
normalized as $\tr t^a t^b=\delta^{ab}/2$. They are hermitian and
traceless, $\tr t^a=0$.

The quark creation and annihilation operators are defined  from
the Fourier mode expansion of the free field operator at light cone
time $x^+=t+z=0$. Since we are focused here on valence quark color charge distributions, we   ignore antiquark contributions to write  (see Appendix II
in~\cite{Brodsky:1989pv}),
\be \label{eq:psi_r_b}
\psi_{i,f}(r)=\int {\dd p^+ \dd^2 p \over {16\pi^3 p^+}} \,
\sum_s \bal{p}{i}{s,f}\, u^s(p)\, e^{-i  p\cdot r} =
\int {\dd x_p \dd^2 p \over {16\pi^3 x_p}} \,
\sum_s \bal{p}{i}{s,f}\, u^s(p)\, e^{-i  p\cdot r}~,
\ee
where $r\equiv (x^+=0,x^-,\vec x_\perp)$ is the coordinate vector. We wrote out spin and flavor indices
explicitly in \eq{eq:psi_r_b} and introduced the momentum fraction $x_p = p^+/P^+$. The integration over
$p^+$ or $x_p$ is restricted to positive values. Using $\bar
u_k\gamma^+ u_p = 2\sqrt{k^+ p^+}$ we can then write
the color charge density operator $\rho^a \equiv j^{+a}$ as
\be
\rho^a(r) = 2 P^+\sum_{\lambda,\lambda'} \int {\dd x_q \dd^2q \over  {16\pi^3\sqrt{x_q}}}
\, \bcl{q}{i}{\lambda} e^{i q\cdot r}
\int {\dd x_p \dd^2p \over {16\pi^3\sqrt{x_p}}}\, \bal{p}{j}{\lambda'}
e^{-i p\cdot r} \, (t^a)_{ij}\, \delta_{\lambda \lambda'}~.\label{densop}
\ee
Note that here $b^\dagger b$ is diagonal in spin and flavor, collectively denoted here by $\lambda$.
Performing a three-dimensional Fourier transform with respect to\ $x^-$ and
$\vec x$, we obtain the color charge density operator in momentum
space,
\be
\trho^a(x_k,\vec k) = \sum_\lambda
\int\limits{\dd x_q \over  \sqrt{x_q(x_q+x_k)}}
 \int \frac{\dd^2q}{16\pi^3}\,
\bcl{q}{i}{\lambda} \bal{k+q}{j}{\lambda} \, (t^a)_{ij}~.
\ee
In this expression, there is a shift of the argument of the
annihilation operator by $(k^+,\vec k)=(x_k
P^+,\vec k)$ relative to the quark creation operator. The physical interpretation of $x_k$ is that it is the longitudinal momentum shift of the quark momentum following an interaction with a colored probe. In the high energy limit, where $P^+$ is large, the $x_k$ dependent corrections are of order $1/P^+$ and can be ignored.  This is explained in Appendix A, where we  show that the density is confined to a thin pancake in $x^-$, with support $1/P^+$. Thus to leading power in $P^+$, we  approximate (in the notation of  \eq{eq:Def_sumint}) $\trho^a(x_k\to0,\vec k) \equiv \trho^a(\vec k) $ so that  
%
\be 
\label{eq:rho^a_k}
\trho^a(\vec k) = \sumint{q,\lambda}{}
\bcl{x_q,\vec q}{i}{\lambda} \, \bal{x_q,\vec k+\vec q}{j}{\lambda} \, (t^a)_{ij} =
\sumint{q,\lambda}{}
\bcl{x_q,\vec q-\vec k}{i}{\lambda}\, \bal{x_q,\vec q}{j}{\lambda} \, (t^a)_{ij}
~.
\ee
The operator in \eq{eq:rho^a_k} differs from that in \eq{eq:number-operator} because there is no shift in the longitudinal momentum.
   We use this expression in the remainder of this paper. Note the variables $(x_q,\vec q)$ are integrated over, so that the left-hand-side depends only on $\vec k$.\\

The
color charge density per unit transverse area, given by the
two-dimensional Fourier transform of this expression, is\footnote{The color charge density is actually given by $\rho^a(\vec x_\perp)$
times the coupling constant  $g$.  However, we prefer to exhibit all factors of
$g$ explicitly and we therefore do not introduce a factor of $g$ in the
definition of $\rho^a(\vec x_\perp)$.}
\be
\rho^a(\vec x_\perp) = \int \frac{\dd^2k}{(2\pi)^2}
\, e^{i\vec k\cdot\vec x_\perp}
\sumint{q,\lambda}{} \,
\bcl{x_q,\vec q-\vec k}{i}{\lambda} \, \bal{x_q,\vec q}{j}{\lambda} \, (t^a)_{ij} ~. \label{rhor}
\ee
In the following subsections, and in the rest of the paper, we will employ an expectation value defined as 
\be
\langle {\bf O}\,\rangle_{K_\perp} = \frac{\bra{P^+,\vec K_\perp} \, {\bf O}\, \ket{P^+, \vec P_\perp=0}}{\langle K|P\rangle}\,,
\label{eq:expectation-value}
\ee
where ${\bf O}$ denotes a generic operator constituted of products of $\rho^a(\vec x_\perp)$ defined above, or its two-dimensional Fourier transform $\trho^a(\vec k)$ in \eq{eq:rho^a_k}. 
The overlap 
$\langle K|P\rangle$ in \eq{eq:ProtonNorm2} is  the standard one given by
\be
\langle K | P\rangle = 16\pi^3 \, P^+\, \delta(K^+-P^+)
\, \delta( \vec{K}_\perp-\vec{P}_\perp) \,.
\ee
We shall be interested in the case when $K^+=P^+$ (see Appendix A). 
\subsection{$\langle\rho^a\rangle$ in the proton}

The proton matrix element of the
color charge density
operator~\eq{eq:rho^a_k}
is given by
\bea
\langle\, \trho^a(\vec q) \, \rangle_{K_\perp} &=&
 \, \tr t^a\,\sum_{\lambda_i} \int \dd x_1\dd x_2 \dd x_3 \, \delta(1-x_1-x_2-x_3) \nonumber\\
&\times &  
\int \frac{\dd^2 p_1 \dd^2 p_2 \dd^2 p_3}{(16\pi^3)^2}
\, \delta(\vec{p}_1+\vec{p}_2+\vec{p}_3)\,
\psi_3^*(k_1,k_2,k_3)\, \psi_3(p_1, p_2, p_3)
~.
\eea
Recall that the arguments of $\psi_3^*$ are given by
$k_i^+= p_i^+\equiv x_iP^+$, $\vec{k}_1 = \vec{p}_1+(1-x_1)\vec{K}_\perp$,
$\vec{k}_2 = \vec{p}_2-x_2\vec{K}_\perp$,
$\vec{k}_3 = \vec{p}_3-x_3\vec{K}_\perp$.

Since $\tr t^a=0$, the above expression is of course zero, as it should
be in QCD. Before we move on to consider higher moments of the charge
charge operator, which are non-zero, it is amusing to consider
what charge conjugation does to the above
expression. $C\, \trho^a(k)\, C^{-1}$ is given
by an expression similar to \eq{eq:rho^a_k} with the replacement
$t^a\to -(t^a)^T= -(t^a)^*$.
Therefore,
\bea
\langle \, C \trho^a(\vec q) \, C^{-1}\, \rangle_{K_\perp} &=& 
- \, \left(\tr t^a\right)^*\,
\sum_{\lambda_i} \int \dd x_1\dd x_2 \dd x_3 \, \delta(1-x_1-x_2-x_3) \nonumber\\
&\times&  
\int \frac{\dd^2 p_1 \dd^2 p_2 \dd^2 p_3}{(16\pi^3)^2}
\, \delta(\vec{p}_1+\vec{p}_2+\vec{p}_3)\,
\psi_3^*(k_1,k_2,k_3)\, \psi_3(p_1, p_2, p_3)~.
\eea
%

\subsection{$\langle\rho^a\rho^b\rangle$ in the proton}

\begin{figure}
  \centering
  \includegraphics[width=0.4\textwidth]{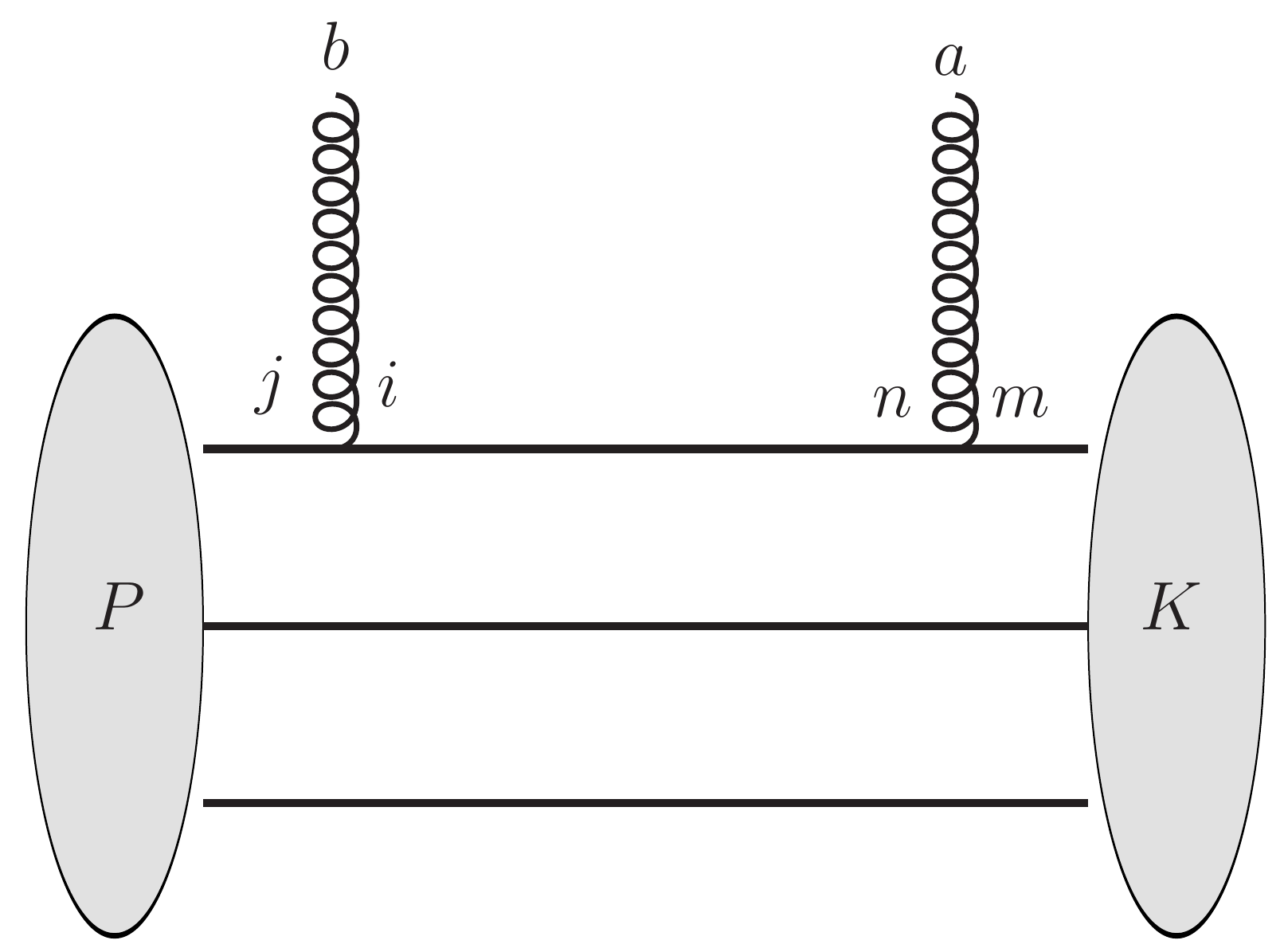}
\hspace{1cm}
   \includegraphics[width=0.4\textwidth]{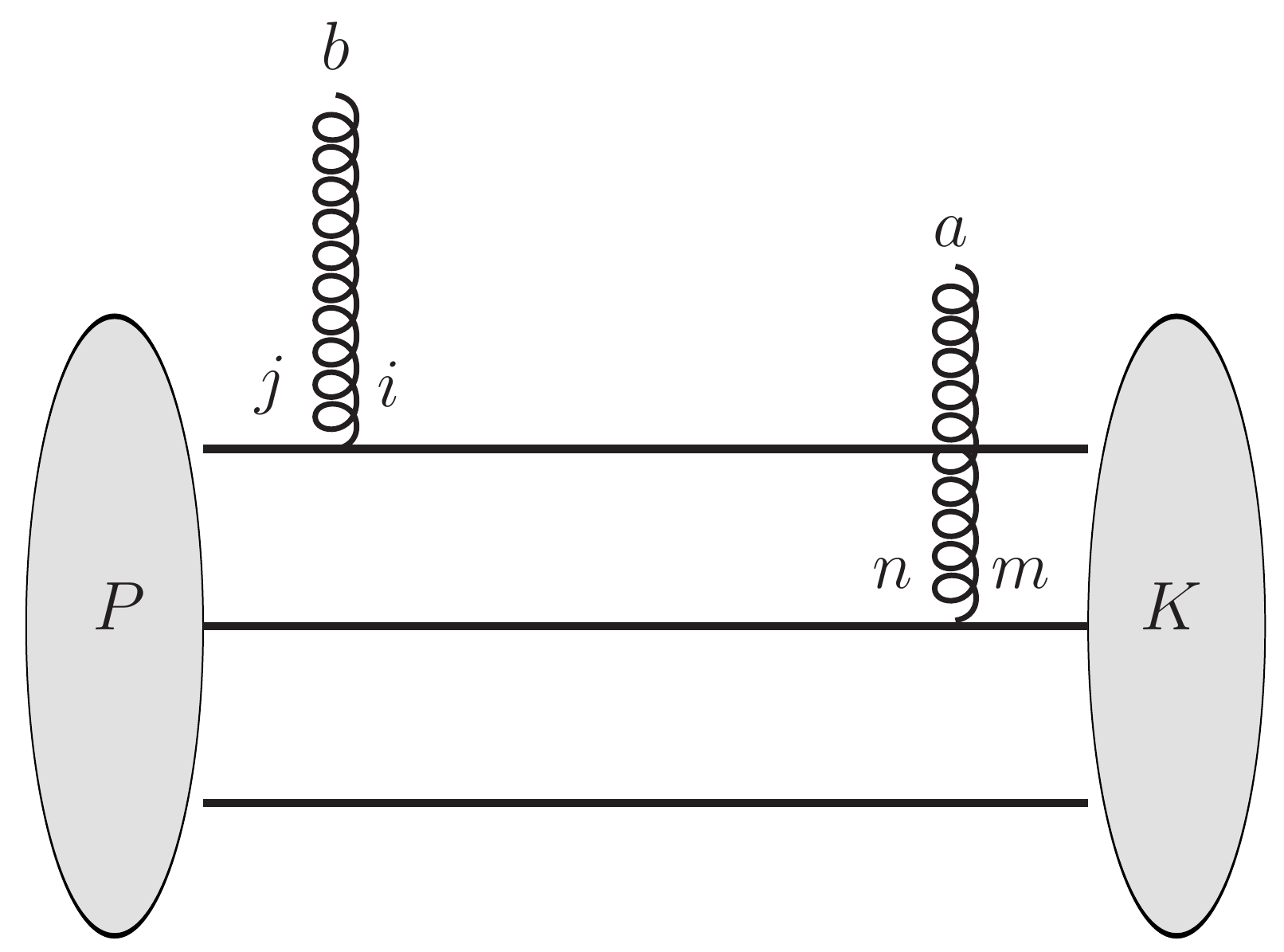}
  \caption{Illustration of the terms we call one-body (left figure)
    and two-body (right figure) contribution to the
    $\langle\rho^a\rho^b\rangle$ correlator. $i, j, n, m = 1,2,3$
    denote the colors of the quarks while $a, b = 1\dots8$ are those
    of the gluons that couple to them.}
\label{fig:one-two-body}
\end{figure}

We shall now compute the first nontrivial color charge correlator, the
expectation value of $\trho^a(q)\, \trho^b(k)$ in the proton. The contributions to its expectation value can be classified, as is common in many-body physics, into one-body and two-body contributions--these are illustrated in Fig.~\ref{fig:one-two-body}. 

We begin with the one-body contribution, where both operators act on the same
quark,
\be
\left[\trho^a(q)\, \trho^b(k)\right]_1 =
\trho^a(q)\, \trho^b(k) \otimes \id \otimes \id
+ \text{permutations}~.
\ee
Then using the anti-commutation relation \eq{anticomm}, and keeping only the one-body contribution leads to:
\be
\bcl{x_{\ell_1},\vec \ell_1-\vec q}{i}{\lambda}\,
\bal{x_{\ell_1},\vec \ell_1}{j}{\lambda}
\bcl{x_{\ell_2},\vec \ell_2-\vec k}{m}{\lambda'}\,
\bal{x_{\ell_2},\vec \ell_2}{n}{\lambda'}
\to
\delcl{\ell_1}{\ell_2-k}{j}{m}{\lambda}{\lambda'}
\,
\bcl{x_{\ell_1},\vec \ell_1-\vec q}{i}{\lambda}\,
\bal{x_{\ell_2},\vec \ell_2}{n}{\lambda'}   ~,
\label{obe}
\ee
and further, using the matrix element of $b^\dagger b$ given previously in
\eq{eq:bc_ba_quark}, we get 
\bea
\epsilon_{i_1 i_2 i_3} \epsilon_{j_1 j_2 j_3}
\bra{p_1',i_1,\lambda_1'; p_2',i_2,\lambda_2'; p_3',i_3,\lambda_3'}
\left[\trho^a(q)\, \trho^b(k)\right]_1
\ket{p_1,j_1,\lambda_1; p_2,j_2,\lambda_2; p_3,j_3,\lambda_3}
&=& \nonumber\\
& & \hspace{-2cm} 3\delta^{ab}
\delc{p_1'}{p_1-q-k}{\lambda_1}{\lambda_1'}
\delc{p_2'}{p_2}{\lambda_2}{\lambda_2'}
\delc{p_3'}{p_3}{\lambda_3}{\lambda_3'}\,.
 \label{eq:rho2_1body}
\eea
 The symmetry of $\psi_3$ under permutations has been used.\\

We will next compute the two-body contributions to the second moment of the color charge
density, where one of the color charge density operators acts on one quark and the other acts on another
quark, as illustrated in Fig.~\ref{fig:one-two-body}. Note that the third quark is a spectator in this process:
\be \label{eq:rho1_rho2_id}
\left[\trho^a(q)\right]_1  \left[\trho^b(k)\right]_2 =
\trho^a(q)\otimes \trho^b(k) \otimes \id + \text{permutations}~.
\ee
Its matrix element is evaluated  to be:
\be
- 3\delta^{ab}\,
\delc{p_1'}{p_1-q}{\lambda_1}{\lambda_1'}
\delc{p_2'}{p_2-k}{\lambda_2}{\lambda_2'}
\delc{p_3'}{p_3}{\lambda_3}{\lambda_3'}
~.  
\label{eq:rho2_2body_disconnected}
\ee
This includes a symmetry factor of 2 and another factor of 3  because there are three such identical terms. 

Summing over both the one-body and two-body terms, the matrix
element of $\trho^a\trho^b$ between Fock states is given by
\bea
& & \epsilon_{i_1 i_2 i_3} \epsilon_{j_1 j_2 j_3}
\bra{p_1',i_1,\lambda_1'; p_2',i_2,\lambda_2'; p_3',i_3,\lambda_3'}
\, \trho^a(q)\, \trho^b(k)\,
\ket{p_1,j_1,\lambda_1; p_2,j_2,\lambda_2; p_3,j_3,\lambda_3}
\nonumber\\
& = &
3\,\delta^{ab}\left\{
\delc{p_1'}{p_1-q-k}{\lambda_1}{\lambda_1'}
\delc{p_2'}{p_2}{\lambda_2}{\lambda_2'}
\delc{p_3'}{p_3}{\lambda_3}{\lambda_3'}
-
\delc{p_1'}{p_1-q}{\lambda_1}{\lambda_1'}
\delc{p_2'}{p_2-k}{\lambda_2}{\lambda_2'}
\delc{p_3'}{p_3}{\lambda_3}{\lambda_3'}
\right\}~.  \label{eq:rho2_Fock_ME}
\eea
As a final step, we need to integrate this expression over the phase space
distribution of the quarks in the proton:
\bea
\langle \, \trho^a(q) \, \trho^b(k) \,\rangle_{K_\perp} &=&\,\frac{1}{2}\,\delta^{ab}\,\sum_{\lambda_i}\int \dd x_1 \dd  x_2 \dd x_3 \, \delta(1-x_1-x_2-x_3) \nonumber\\
&\times&  \int {\dd^2 p_1 \dd^2 p_2 \dd^2 p_3\over (16\pi^3)^2}
\, \delta(\vec{p}_1+\vec{p}_2+\vec{p}_3) \left[\psi_3^*(k_1,k_2,k_3)-\psi_3^*(\bar k_1,\bar k_2,\bar k_3) \right]
\psi_3(p_1, p_2, p_3)~.\nonumber\\
\label{eq:rho2_Kt}
\eea
The arguments of $\psi_3^*$ are $k_i^+=\bar k_i^+=x_i P^+$, $\vec
k_1=\vec p_1 + (1-x_1) \vec K_\perp$, $\vec k_2=\vec p_2 -x_2 \vec
K_\perp$, $\vec k_3=\vec {\bar k}_3=\vec p_3 -x_3 \vec K_\perp$, $\vec
{\bar k}_1=\vec p_1 -\vec q-x_1\vec K_\perp$, $\vec {\bar k}_2=\vec
p_2 -\vec k -x_2 \vec K_\perp$, and all flavors and helicities with $\lambda_i'=\lambda_i$.
Note that the r.h.s.\ does depend on $\vec q$ and $\vec k$, even at fixed
momentum transfer $\vec{K}_\perp$, because $\vec{\bar k}_1$ and $\vec
{\bar k}_2$ depend on $\vec q$, $\vec k$. The factor in brackets is  a momentum conserving delta function, arising from the normalization of plane-wave states, that does not appear in invariant amplitudes. The factor in parentheses results from the color algebra. The remaining term is a color-charge form factor, $\cG$ that contains  intrinsically non-perturbative information on the
color charge distributions in the three valence quark state of the
proton. Thus we rewrite \eq{eq:rho2_Kt} as
\bea
\langle\, \trho^a(q) \, \trho^b(k) \,\rangle_{\vec K_\perp} &=&
\frac{1}{2}\,\delta^{ab}\,{\cal G}(\vec k,\vec K_\perp)~,
\label{eq:Pom-FF1}
\eea
with 
\bea & {\cal G}(\vec k,\vec K_\perp)\equiv  \cG_1(\vec K_\perp)- \cG_2(\vec k,\vec K_\perp)\label{cgdef}\\
& \cG_1(\vec K_\perp)=
 \int \dd \mathrm{PS}_3\,\psi_3^*(p_1+(1-x_1)\vec K_\perp,p_2-x_2\vec K_\perp,p_3-x_3\vec K_\perp)  
\psi_3(p_1, p_2, p_3)\\
&\cG_2(\vec k,\vec K_\perp)=
 \int \dd \mathrm{PS}_3\,\psi_3^*(p_1+\vec k+(1-x_1)\vec K_\perp,p_2-\vec k -x_2\vec K_\perp,p_3-x_3\vec K_\perp ) 
\psi_3(p_1, p_2, p_3)~. \ 
\label{eq:G2}
\eea
The hybrid notation $p_1+(1-x_1)\vec K_\perp $ {\it etc.} means that the quantum numbers of $p_1$ are unchanged, except that the transverse momentum is increased by $(1-x_1)\vec K_\perp$. 
Note further that $ \mathrm{PS}_3$ is a compact notation for the sum over helicities and momentum phase space integrals in  \eq{eq:rho2_Kt}.\\

The form factor $\cG$ enters in calculations of the two-gluon exchange
model of the Pomeron \cite{Gunion:1976iy}.  Those early authors used
simple models in their evaluations. The present formulation is more
general and allows for the inclusion of a variety of models; see for
example~\cite{Schlumpf:1992vq,Frank:1995pv,Miller:2002ig,Pasquini:2007iz,Pasquini:2009bv,Lorce:2011dv,Cloet:2012cy}.\\

For forward scattering, $\vec K_\perp=0$, 
\bea \cG(\vec k,0)=1-\cG_2(\vec k,0)\,.
\label{eq:color-neutral}\eea
This quantity vanishes as $|\vec k|$ approaches 0, because $\cG_2(0,0)=1$, according to the normalization condition for $\psi_3.$ This  vanishing of $\cG(\vec k,0)$, caused by the influence of color neutrality,  leads to the suppression of infrared divergences.\\

\subsection{Relation to the  McLerran-Venugopalan (MV)  model}
  
It is worthwhile  and interesting to compare our results for the proton with those of  the  MV
model~\cite{McLerran:1993ni,McLerran:1993ka,McLerran:1994vd}
approximation, valid for a large nucleus of radius $R$. In the first MV paper~\cite{McLerran:1993ni}, $\mu^2 $ is defined by the relation
\bea 
\la \rho(\vec x_\perp)\rho(\vec y_\perp)\ra_{K_\perp=0}= \mu^2\,
\delta(\vec x_\perp-\vec y_\perp),
\eea
where $\mu^2$ is the average square of the color charge per unit area. In the original MV model, only the case of zero momentum transfer $K_\perp=0$ between the initial and final states of the 
nucleus was considered. Since $\rho$ has dimensions of inverse area,
the state defined by the brackets must have  no dimensions.\\

Later work (see for example \cite{Krasnitz:2002mn}) showed that
$\mu^2$ is a function that can depend on $\xp,\yp$ and the expression
above can be generalized to
\bea \int d^2R_\perp \la
\rho^a(\vec R_\perp+\vec s_\perp/2)\, \rho^b(\vec R_\perp-\vec
s_\perp/2)\ra_{K_\perp=0} = \delta^{ab}\, \mu^2_{\rm
  MV}(\vec s_\perp)\label{newdef} ~.\eea

Our formulation is in terms of momentum, so here we take the state
$|\cdots\ra$ to be the momentum eigenstate $|P\ra$ and Fourier
transform by operating with $\int d^2s_\perp e^{-i\vec
  k_\perp\cdot\vec s_\perp}$ on both
sides of \eq{newdef}.  The result is
\bea \delta^{ab} \int d^2s_\perp e^{-i\vec k_\perp\cdot\vec s_\perp}
\,
\mu^2_{\rm MV}(\vec s_\perp)\equiv
\widetilde{\mu^2}_{\rm MV}(\vec k_\perp)=\la
\rho^a(\vec k_\perp)\, \rho^b(-\vec k_\perp)\ra_{K_\perp=0}\,.
\eea
As suggested previously~\cite{Gavai:1996vu,Lam:1999wu}, and as shown explicitly in \cite{Krasnitz:2002mn}, imposing a color neutrality condition $\int d^2 x_\perp \rho^a(x_\perp)=0$ over a radial distance of $1/\Lambda$, where $\Lambda$ is a color neutralization scale, gives
\be
\widetilde{\mu^2}_{\rm MV}(\vec k_\perp) \rightarrow 0 \,\,\,{\rm
  for}\,\,\,
\kp\rightarrow 0\,, \,\,\,{\rm and}\,\,\,
\widetilde{\mu^2}_{\rm MV}(\vec k_\perp) = {\rm constant}\,\,\, {\rm for}\,\,\, \kp > \Lambda\, .
\label{eq:KNV}
\ee

In the approach employed here, the use of \eq{cgdef}, and the dimensionless momentum eigenstate leads to the result:
\bea
\widetilde{\mu^2}_{\rm MV}(\vec k_\perp)= {(N_c^2-1)\over 2}
\left(1-\cG_2(\vec k_\perp)\right).\label{goody}
 \eea
Just as in \eq{eq:KNV}, based on the normalization constraint on
$\cG_2(\vec k_\perp)$ discussed after \eq{eq:color-neutral},
$\widetilde{\mu^2}_{\rm MV}(\vec k_\perp)$ vanishes for $\kp\to0$. The
structure of $\cG_2(\vec k_\perp)$ in the $K_\perp=0$ limit of \eq{eq:G2}
suggests on general grounds that it vanishes at large values of
$k_\perp$.  The latter limit corresponds to the MV
model~\cite{McLerran:1993ni,McLerran:1993ka,McLerran:1994vd}
approximation, valid for a large nucleus:
 \be \frac{1}{2}\cG(\vec q,0)\to\bar{ \mu^2}_{\rm
  MV}\,\Theta(q^2-\Lambda^2)\,.
\label{approx}
\ee
Relating  \eq{goody} to  \eq{approx} allows the  identification of  the scale $\Lambda$ with a momentum on the order of the inverse of the radius of the proton.\\

We can apply the formalism computed thus far to compute the gluon
distribution of the
proton~\cite{Mueller:1999yb,Iancu:2000hn,Iancu:2003xm}.  The number of
gluons in the hadron wavefunction, having longitudinal momenta between
$\xx P^+$ and $(\xx+d\xx)P^+$, and a transverse size $\Delta x_\perp
\sim 1/Q\,$, is denoted as $G(\xx,Q^2) d\xx$, and is given by
\be
\label{eq:GDF}
\xx G(\xx,Q^2)=\frac{1}{\pi}\int {d^2k_\perp \over (2 \pi)^2}\,\Theta(Q^2-
k_\perp^2)\bigl\langle F^{i+}_a(\vec k_\perp)
F^{i+}_a(-\vec k_\perp)\bigr\rangle_{K_\perp=0}\,,
\ee
where $ F^{i+}_a$ is the color-electric field.\\

Solving the Yang-Mills equations in light cone gauge, to linear order in the color charge density, one obtains~\cite{Iancu:2003xm}
\be
{F}^{+i}_a(\vec k_\perp)
\,\simeq\,i g{k^i \over k_{\perp}^2}\,
{\rho^a(\vec k_\perp)}\,,
\label{aaimom}
\ee
and 
\be\label{linphi}
\langle{ F}^{i+}_a(\vec k_\perp){F}^{i+}_a(-\vec k_\perp)\rangle_{K_\perp=0}
\simeq\frac{g^2}{ k_{\perp}^2}\,\langle\rho^a(\vec k_\perp)\,
\rho^a(-\vec k_\perp)\rangle_{K_\perp=0}\,. 
\ee
Inserting this expression in \eq{eq:GDF} and using \eq{cgdef}
one obtains the expression:
\bea
\label{MV-LOW}
 \xx G(\xx,Q^2)&\simeq&\frac{g^2}{4\pi^2}\, {(N_c^2-1)\over2}\int^{Q^2}_{0}
\frac{dk_\perp^2}{ k_{\perp}^2}(1-\cG_2(\vec k_\perp,0)) \,.
\label{xgn}
\eea
A comparison of \eq{xgn} with the corresponding expression in
\cite{Iancu:2001md} reaffirms the result in \eq{goody}.  Note that the
integral over $k_{\perp}$ does not have an infrared divergence. As
discussed earlier, this is a consequence of the color neutrality of
the nucleon. If one breaks up the integral in \eq{xgn} into a piece
from $0\ < k_\perp < \Lambda$ and another from $\Lambda < k_\perp <
Q$, the former will integrate to a constant while the latter will give
a factor $\frac{\alpha_S N_c}{\pi} C_F\ln(Q^2/\Lambda^2)$, where
$\alpha_S=g^2/4\pi$ and $C_F= (N_c^2-1)/2N_c$ is the Casimir of a
quark in the fundamental representation. Thus in the Bjorken limit of
$Q^2\rightarrow \infty$, one obtains the usual leading
contribution~\cite{Kovchegov:2012mbw} to the gluon distribution
\be
\xx G(\xx,Q^2)\approx \frac{\alpha_S N_c}{\pi} C_F\ln(Q^2/\Lambda^2)\,. 
\ee

Interestingly, the effect of color neutralization as imposed on the MV
model is also obtained by QCD evolution of the MV model to small
$x$~\cite{Iancu:2001md,Mueller:2001uk}. Gluons emitted by the quarks
screen each other at a saturation scale
$Q_S(x)$~\cite{Gribov:1984tu,Mueller:1985wy}; for small $x$,
$Q_S^2(x)\gg \Lambda^2$.  More specifically, $Q_S^2\propto\mu_{\rm
  JIMWLK}^2$, where $\mu_{\rm JIMWLK}^2$ is the variance of the
Gaussian weight functional for $W[\rho]$ that reproduces the
Balitsky-JIMWLK hierarchy~\cite{Balitsky:1995ub,Weigert:2000gi} in the
CGC EFT. However, while numerical simulations suggest that there is a
renormalization group (RG) flow to this Gaussian fixed
point~\cite{Dumitru:2011vk}, it remains an open question at what
values of $x$ this is achieved. This concern is in particular germane
to the proton, where the color charge densities are not {\it a priori}
large. \\

Nevertheless, even if the Gaussian approximation of the CGC EFT is not
robust, one can still make considerable progress by computing $\langle
\trho\trho\rangle$ from first principles on the light front. Even
though our result for $\langle \trho\trho\rangle$ is for the three
valence quark state, it is straightforward, with some effort, to
extend it to include Fock states containing gluons. A more important
issue though is that higher combinants
$\langle\trho^a(q_1)\trho^b(q_2)\cdots \trho^k(q_n)\rangle$ for $n\geq
3$ cannot be expressed in terms of $\langle \trho\trho\rangle$, as
they would be if $W[\rho]$ had a Gaussian form. \\

In our approach, these higher combinants can be computed without invoking a $W[\rho]$ functional at all! These can be
computed explicitly and expressed in terms of the corresponding color
charge form factors, as in \eq{cgdef}. The latter, as we shall
illustrate in subsequent sections, can be extracted
from exclusive measurements in DIS at large $x$. Besides our intrinsic
interest in the shape and momentum distribution of color charges at
large $x$, an important consequence, for the RG
discussion above, is a novel strategy whereby  one can study systematically the many-body RG
flow of these color charge distributions to the putative Gaussian
fixed point. To illustrate this strategy, we will compute
$\langle\rho^a\rho^b\rho^c\rangle$ for the three quark valence state
and identify the corresponding color charge form factor. This will also have interesting consequences in its own right, which we shall discuss in Section IV. 

\subsection{$\langle\trho^a\trho^b\trho^c\rangle$ in the proton}

\begin{figure}
  \centering
   \includegraphics[width=0.45\textwidth]{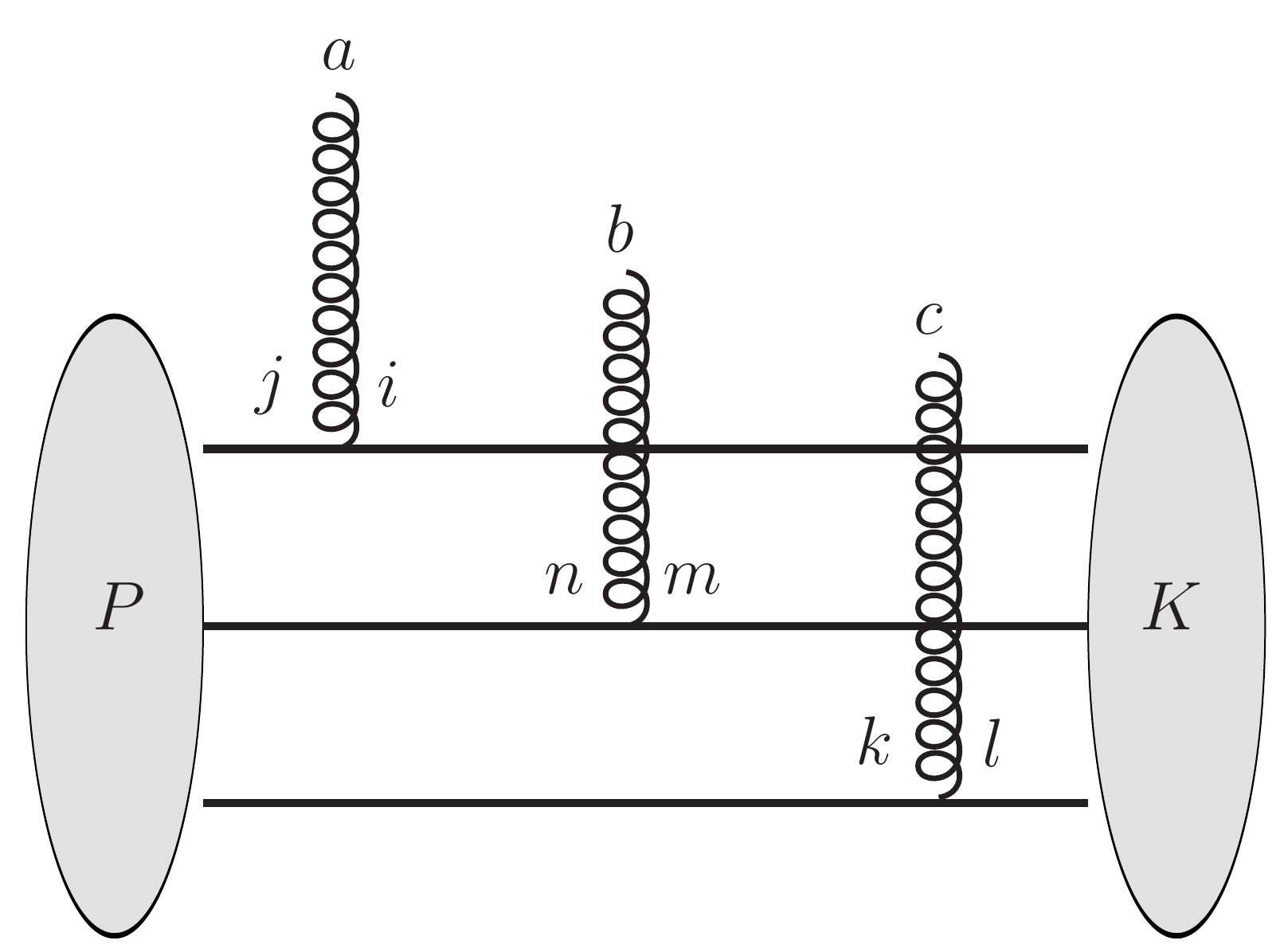}
  \caption{Illustration of the three-body contribution to the $\langle\rho^a\rho^b\rho^c\rangle$ correlator.}
\label{fig:rho-rho-ThreeBody}
\end{figure}

To compute the expectation value of $\trho^a(q_1)\, \trho^b(q_2)\,
\trho^c(q_3)$ in the proton, in addition to the one-body and two-body
terms discussed previously, we will have an additional three-body
term, which is illustrated in Fig.~\ref{fig:rho-rho-ThreeBody}.

\subsubsection{One-body contribution}

As previously for $\langle \trho\trho\rangle$, we start with the
one-body contribution where all three charge
operators act on the same quark.
Defining this term as
\be
   [\trho^a(q_1)\, \trho^b(q_2)\, \trho^c(q_3)]_1 =
   \trho^a(q_1) \trho^b(q_2)\trho^c(q_3) \otimes \id \otimes \id
   + \text{permutations}~,
\ee
we find
\bea
& &\langle\, [\trho^a(q_1) \, \trho^b(q_2) \, \trho^c(q_3)]_1
\,\rangle_{K_\perp} = \, \tr t^at^bt^c  \int \dd x_1\dd x_2 \dd x_3 \, \delta(1-x_1-x_2-x_3)  \nonumber\\
& \times& 
\int {\dd^2 p_1 \dd^2 p_2 \dd^2 p_3\over (16\pi^3)^2}
\, \delta(\vec{p}_1+\vec{p}_2+\vec{p}_3) \sum_{\lambda_i}
\psi_3^*(k_1,k_2,k_3)\, \psi_3(p_1, p_2, p_3)~.\nonumber \\
\label{eq:rho3_1body_Kt}
\eea
The arguments of $\psi_3^*$ are $k_i^+=x_i P^+$, $\vec k_1=\vec p_1 +
(1-x_1) \vec K_\perp$, $\vec k_2=\vec p_2 -x_2 \vec K_\perp$, $\vec
k_3=\vec p_3 -x_3 \vec K_\perp$, and all flavors and helicities
unchanged ($\lambda_i'=\lambda_i$).
The color factor is given by
\be
\tr\, t^a t^b t^c = \frac{1}{4}d^{abc} + \frac{i}{4}f^{abc}.
\ee
 Since the $\vec k_i$ do not
explicitly involve the $\vec q_i$, it follows that at fixed $\vec
K_\perp$, the expectation value of the one-body term is a constant times the delta function constraint on their momentum arguments.  

\subsubsection{Two-body contribution}

The computation of the two-body contribution follows analogously to
previously. In this case, two of the charge operators act on one
quark, while the third $\rho$-operator acts on a second quark.  There are three separate terms, corresponding to the three different possible spectator quarks.
The first term can be written as 
\bea
\left[\trho^a(q_1)\, \trho^b(q_2)\right]_1\, \left[ \trho^c(q_3)\right]_2
&=&
\trho^a(q_1)\, \trho^b(q_2)
\otimes
\trho^c(q_3)
\otimes \id  + \text{permutations}~.  \label{eq:rho3_2body_first}
\eea
We then find,
\bea
&&\langle \,[ \trho^a(q_1) \, \trho^b(q_2)]_1 \,[ \trho^c(q_3)]_2
\,\rangle_{K_\perp} = 
-\tr\, t^at^bt^c \,  \int \dd x_1\dd x_2 \dd x_3 \, \delta(1-x_1-x_2-x_3)  \nonumber\\
&\times&
\int {\dd^2 p_1 \dd^2 p_2 \dd^2 p_3\over (16\pi^3)^2}
\, \delta(\vec{p}_1+\vec{p}_2+\vec{p}_3) \sum_{\lambda_i}
\psi_3^*(k_1,k_2,k_3)\, \psi_3(p_1, p_2, p_3)~.
\label{eq:rho3_2body_Kt_part1}
\eea
Here $k_i^+=x_i P^+$, $\vec k_1=\vec p_1 + \vec q_3 + (1-x_1)\vec
K_\perp$, $\vec k_2=\vec p_2 -\vec q_3 - x_2 \vec K_\perp$, $\vec
k_3=\vec p_3 - x_3 \vec K_\perp$. As usual, all flavors and helicities
are kept unchanged ($\lambda_i'=\lambda_i$). For the other two
two-body contributions, one needs to exchange $\vec q_3$ in $\vec k_1$ and
$\vec k_2$ by $\vec q_1$ and $\vec q_2$, respectively. Moreover, the
  color factor for the expectation value of $[\trho^a(q_1) \,
    \trho^c(q_3)]_1 \,[ \trho^b(q_2)]_2$ is $\tr t^at^ct^b$ instead of
  $\tr t^at^bt^c$.

Unlike the one-body contributions, these contributions do depend on
$\vec q_i$, even at fixed $t=-K^2_\perp$. Note that if one writes
$\vec q=\vec q_1 + \vec q_2 = -\vec q_3-\vec K_\perp$ and $\vec k =\vec q_3$,
the phase space integral in \eq{eq:rho3_2body_Kt_part1} is identical
to the one which appeared in the two-body contribution to
$\left<\trho^a(q)\trho^b(k)\right>$ in \eq{eq:rho2_Kt}.  This identity
can be seen by direct comparison and serves as a check on the
computation. 

\subsubsection{Three-body contribution}

The three-body operator corresponds to each color charge operator
acting on separate valence quarks--see
Fig.~\ref{fig:rho-rho-ThreeBody}. Defining this term as
\be
[\trho^a(q_1)\trho^b(q_2) \trho^c(q_3)]_3
=
\trho^a(q_1)\otimes \trho^b(q_2)
\otimes
\trho^c(q_3)
+ \text{permutations}~,
\ee
we find 
\bea
& &\langle \, [\trho^a(q_1) \, \trho^b(q_2) \, \trho^c(q_3)]_3
\,\rangle_{K_\perp} = \frac{1}{2}d^{abc} \, \int \dd x_1\dd x_2 \dd x_3 \, \delta(1-x_1-x_2-x_3)
 \nonumber\\
& \times&  
\int {\dd^2 p_1 \dd^2 p_2 \dd^2 p_3\over (16\pi^3)^2}\,
\delta(\vec{p}_1+\vec{p}_2+\vec{p}_3) \sum_{\lambda_i}
\psi_3^*(k_1,k_2,k_3)\, \psi_3(p_1, p_2, p_3)~.
\label{eq:rho3_3body_Kt}
\eea
Here, $k_i^+=x_i P^+$, $\vec k_1=\vec p_1 -\vec q_1 -x_1 \vec
K_\perp$, $\vec k_2=\vec p_2 -\vec q_2 - x_2 \vec K_\perp$, $\vec
k_3=\vec p_3 -\vec q_3 - x_3 \vec K_\perp$. As usual, all flavors and
helicities are unchanged ($\lambda_i'=\lambda_i$). As in the two-body case, this three-body contribution 
depends on $\vec q_i$, even at fixed $t=-K^2_\perp$.\\

Our net result for $\langle \trho^a (q_1)\trho^b (q_2) \trho^c
(q_3)\rangle$ is the sum of \eq{eq:rho3_1body_Kt},
\eq{eq:rho3_2body_Kt_part1} (plus the permutations of momenta
indicated below that equation), and \eq{eq:rho3_3body_Kt}. Both the
symmetric and antisymmetric structure factors, respectively $d^{abc}$
and $f^{abc}$, are proportional to color charge form
factors. Specifically, we can express the symmetric (S) piece as\footnote{We
introduce an explicit factor of $1/N_c$ on the right hand side in
order to match powers of $N_c$ in the odderon amplitude to a
computation in perturbative QCD~\cite{Kovchegov:2003dm}, see below.}
\bea
\la [\trho^a(q_1) \trho^b(q_2)\trho^c(q_3)]_{\rm S}\ra_{K_\perp}
&\equiv& \frac{d^{abc}}{N_c}\,\, \cG_O(\vec q_1,\vec q_2,\vec q_3 ; \vec K_\perp)\,,
\label{eq:Odderon-form-factor}
\eea
which involves the 1, 2 and 3-body terms. Anticipating results to
appear, we denote $\cG_O$ to be the Odderon form factor.\\

We note that similar form factors were discussed previously in the
context of high energy forward scattering
amplitudes~\cite{FukugitaKwiecinski,Czyzewski:1996bv}.  Fukugita and
Kwiecinski~\cite{FukugitaKwiecinski} similarly identified one-body,
two-body and three-body contributions and noted that the two-body
contribution can be expressed in terms of the Pomeron form factor in
\eq{cgdef}. However, though they suggest that the three-body
contribution in \eq{eq:rho3_3body_Kt} can be expressed in terms of the
two body contribution, our results show that this is not true in
general. Furthermore, unlike these works, we are able to express our
results explicitly in terms of the QCD valence Fock state
wavefunction.\\

We can however confirm the observation in \cite{Czyzewski:1996bv} that
in the limit that any of the $\vec q_i\to 0$, the sum of all these
contributions should vanish. Specifically, taking $\vec q_3\to0$ (but
$\vec q_1$, $\vec q_2$, $\vec K_\perp$ arbitrary!), one observes that
the sum of the $a$, $b$, $c$-symmetric pieces of
\eq{eq:rho3_1body_Kt}, \eq{eq:rho3_2body_Kt_part1} and
\eq{eq:rho3_3body_Kt} does indeed vanish. The underlying reason is a
general feature of QCD that must be satisfied by any model: a long
wavelength gluon cannot couple to a color singlet.

\section{Color charge form factors and exclusive heavy quark production in DIS}

In the previous section, we derived explicit expressions for the
expectation values of quadratic and cubic combinants of the color
charge density and reexpressed the results in terms of nonperturbative
color charge form factors. We show here that these
nonperturbative quantities can be determined from exclusive
measurements of heavy Quarkonia in DIS at large $x$ at Jefferson
Laboratory~\cite{Hafidi:2017bsg,Joosten:2018gyo,Joosten:2018fql} and
in the future at an Electron-Ion Collider~\cite{Accardi:2012qut}.  We derive the 
 amplitude  for exclusive
quarkonium production and express it in terms our Pomeron
and Odderon color charge form factors in the   first subsection. Specifically, we show that the exclusive $J/\Psi$ cross-section is proportional to 
both the Pomeron and Odderon form factors. In contrast, 
the  $\eta_c$ amplitude depends only on the Odderon form factor; the latter  can therefore 
be extracted directly from an exclusive measurement of the production of $\eta_c$ mesons. While
this possibility is well known in the literature, and even discussed
very recently~\cite{Goncalves:2015hra}, we  will articulate how our
work brings a novel perspective to this discussion.

\subsection{Amplitude for exclusive quarkonium production at large $x_{\rm Bj}$}

In DIS at high energies, the amplitude for exclusive quarkonium
production be expressed as~\cite{Kowalski:2006hc}
\be 
{\cal A}^{\gamma^*\, p\to Q{\bar Q}\, p} (Q^2,\vKT)\sim i \int \dd^2r
\int_0^1 \frac{dz}{4\pi}
\, \left(\Psi_{\gamma^*}
\Psi^*_{Q{\bar Q}}\right)(\vec r,z,Q^2) ~ e^{-i\frac{(1-2z)}{2}\vec r\cdot \vKT}
\int d^2 b_\perp \, e^{i {\vec b}_\perp\cdot \vKT}~{\cal T}(\vec
r,\vec b_\perp; \vec K_\perp)\, .
\label{eq:Onium-amplitude}
\ee
Here $\Psi_{\gamma^*}$ is the light cone wave function of a virtual
photon to fluctuate into a charm-anticharm pair~\cite{Nikolaev:1990ja}
of relative size $\vec r$, $z$ ($1-z$) is the fraction of the photon
momentum taken by the quark (antiquark) and $\vKT$ is the transverse momentum transfer between the incoming and outgoing proton. Further,
$\Psi_{Q{\bar Q}}(\vec r,z,Q^2)$ is the wavefunction  corresponding to the
overlap $\langle c{\bar c}|Q{\bar Q}\rangle$ of the $c{\bar c}$ pair
with any $Q{\bar Q}$ quarkonium state ($J/\Psi$, $\Psi(2S)$, $\eta_c$,
$\chi_c,\cdots$). 

\begin{figure}
  \centering
   \includegraphics[width=0.45\textwidth]{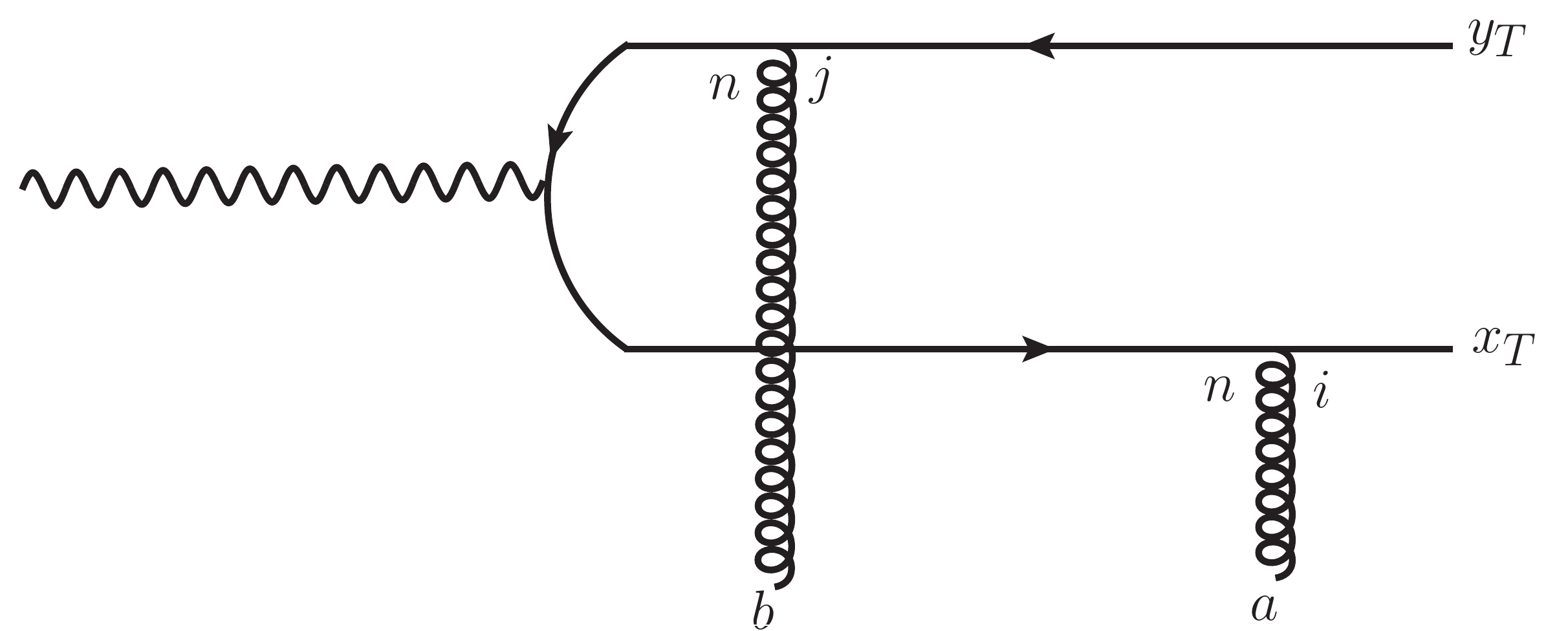}
  \caption{Illustration of the two-gluon contribution to the DIS exclusive amplitude for $c{\bar c}$ production.}
\label{fig:two-gluon}
\end{figure}

Finally, ${\cal T}$ denotes the invariant amplitude for elastic scattering
of the $c{\bar c}$ pair off color fields in the target
proton\footnote{We use the shorthands $\int\limits_{x_T} \equiv \int
  \dd^2x_T$ while $\int\limits_{q} \equiv \int
  \frac{\dd^2q}{(2\pi)^2}$.} and can be expressed
as\footnote{In \cite{Kowalski:2006hc}, the factor of $N_c$ is absorbed in the definitions of $\Psi_{\gamma^*}$  and $\Psi_{Q{\bar Q}}(r,z,Q^2)$; we feel it is more appropriate to not do so and to keep it explicit in ${\cal T}$. To avoid double counting, this should be taken into account while using \eq{eq:Onium-amplitude}.}~\cite{Dominguez:2011wm,Hatta:2017cte,Roy:2018jxq} :
\be 
   {\cal T} (\vec r,\vec b_\perp; \vec K_\perp) = 2\, N_c\,
   \left[ 1 - \frac{1}{N_c}\,{\rm tr} \,\la U\left(\vec b_\perp +
     \frac{\vec r}{2}\right)
     U^\dagger\left( \vec b_\perp - \frac{\vec r}{2}\right)\ra_{\vec
       K_\perp} \right]\, .
\label{eq:Mij(ell,P)}
\ee
Here $U$ (and $U^\dagger$) are lightlike Wilson lines representing the
color rotation of a color dipole in the gauge field background of the
proton. The brackets $\langle\cdots \rangle_{K_\perp}$ represent
taking the expectation value in the proton according to
\eq{eq:expectation-value}.  As in the discussion there, and discussed
further in Appendix A, we are making an eikonal approximation that the
proton target has a large $P^+$ momentum. In writing
\eq{eq:Mij(ell,P)}, we identified the coordinates $\vec x_T$ and $\vec
y_T$ of the quark-antiquark pair shown in Fig.~\ref{fig:two-gluon}
with the impact parameter of the quark-antiquark pair and their
relative separation respectively as~\cite{Bartels:2003yj},
\bea
{\vec {\tilde b}_\perp} &=& z {\vec x}_\perp + (1-z) {\vec y}_\perp \,,\nonumber\\
{\vec r} &=& {\vec x}_\perp - {\vec y}_\perp \, ,
\eea
and then a further transformation~\cite{Hatta:2017cte}
\be
{\vec b}_\perp = {\vec {\tilde b}_\perp} + \left(\frac{1}{2} -z\right)\,\vec{r}\, ,\
\ee
to express the result in the symmetric form shown in
\eq{eq:Mij(ell,P)}. The phase factor $e^{-i\frac{(1-2z)}{2}\vec r\cdot
  \vKT}$ in \eq{eq:Onium-amplitude} is a consequence of these
transformations.

In Lorenz gauge $\partial_\mu A^\mu=0$, and in the above described
eikonal approximation, the gauge fields appearing in the Wilson lines
corresponding to multiple scattering of a quark at spatial position
$(x^-,\vec x_T)$ have only one component $A^+$, which satisfies the Poisson
equation $\nabla_\perp^2 A^+ = g\rho(x^-,\vec x_\perp)\,$ ($A_\mu\equiv
t^aA^a_\mu$) and the lightlike Wilson lines are path ordered in the
$x^-$ direction~\cite{Iancu:2003xm,Kovchegov:2012mbw}:
\bea
U^\dagger(\vec x_T) &=& {\cal P} e^{ig \int dx^- A^+(x^-,\vec x_T)} \\
&=& 1 + ig \int dx^- A^+(x^-,\vec x_T) +
(ig)^2 \int dx^- \int^{x^-} dy^- A^+(x^-,\vec x_T) A^+(y^-,\vec x_T)
\nonumber\\
&+ &
(ig)^3 \int dx^- \int^{x^-} dy^- \int^{y^-} dz^- A^+(x^-,\vec x_T)
A^+(y^-,\vec x_T) A^+(z^-,\vec x_T) + \cdots ~.
\eea
where the factors of $g\,t^a$ contained in this expansion correspond to the vertices arising from the order by order expansion of the 
coherent coupling of the gluon fields in the target to the $c$ or $\bar c$ quark.\\

Expanding $U(\vec x_T) U^\dagger(\vec y_T)-1$ to third order in
$gA^+$  gives:
\bea
1-U(\vec x_T) U^\dagger(\vec y_T)  &=&
 (ig)^2 \int dx^- \int dy^- A^+(x^-,\vec x_T) A^+(y^-,\vec y_T) \nonumber\\
& & -(-ig)^2 \int dx^- \int_{x^-} dy^- A^+(x^-,\vec x_T) A^+(y^-,\vec x_T) \nonumber\\
& & -(ig)^2 \int dx^- \int^{x^-} dy^- A^+(x^-,\vec y_T) A^+(y^-,\vec y_T) \nonumber\\
& & - (-ig)(ig)^2 \int dx^- \int dy^- \int^{y^-} dz^- A^+(x^-,\vec x_T)
A^+(y^-,\vec y_T) A^+(z^-,\vec y_T)\nonumber\\
& & - (-ig)^2(ig) \int dx^- \int_{x^-} dy^- \int dz^- A^+(x^-,\vec x_T)
A^+(y^-,\vec x_T) A^+(z^-,\vec y_T)\nonumber\\
& & - (-ig)^3 \int dx^- \int_{x^-} dy^- \int_{y^-} dz^- A^+(x^-,\vec x_T)
A^+(y^-,\vec x_T) A^+(z^-,\vec x_T) \nonumber\\
& & - (ig)^3 \int dx^- \int^{x^-} dy^- \int^{y^-} dz^- A^+(x^-,\vec y_T)
A^+(y^-,\vec y_T) A^+(z^-,\vec y_T) + \cdots ~.
\label{eq:UxU+y-1}
\eea

Let us first consider the expectation value of the previous expression up to
order $(gA^+)^2$. Using the fact that it is symmetric under $x^- - y^- \to y^- -
x^-$, we can express the term appearing in \eq{eq:Mij(ell,P)} as 
\bea 
1- \la U\left(\vec b_\perp + \frac{\vec r}{2}\right)
U^\dagger\left( \vec b_\perp - \frac{\vec r}{2}\right) \ra_{K_\perp}
&=&
- g^2 \int dx^-
\int dy^- \la A^+\left(x^-,\vec b_\perp + \frac{\vec r}{2}\right)
A^+\left(y^-,\vec b_\perp - \frac{\vec r}{2}\right)\ra_{K_\perp}
\nonumber\\
&+&  \frac{1}{2} g^2 \int dx^- \int dy^-
\la A^+\left(x^-,\vec b_\perp + \frac{\vec r}{2}\right)
A^+\left(y^-,\vec b_\perp + \frac{\vec r}{2}\right)\ra_{K_\perp} \nonumber\\ 
&+ & 
\frac{1}{2} g^2 \int dx^- \int dy^- \la A^+\left(x^-,\vec b_\perp - \frac{\vec r}{2}\right)
A^+\left(y^-,\vec b_\perp - \frac{\vec r}{2}\right)\ra_{K_\perp} ~.
\eea
We can use the Poisson equation to relate $A^+$ to the charge density
operator $\rho$ and further, to write the latter in terms of its
two-dimensional Fourier representation. In doing so, note that the
integral of $\trho(x^-,\vec q)$ over $x^-$ corresponds to the operator
$\trho(q)$ in \eq{eq:rho^a_k}. We then obtain, to quadratic order in
$A^+$ or $\rho$,
\bea
 1-\frac{1}{N_c} {\rm tr} \la U\left(\vec b_\perp + \frac{\vec
   r}{2})\right)
 U^\dagger\left(\vec b_\perp - \frac{\vec r}{2}\right)\ra_{\vec K_\perp}^{O(\rho^2)}
&= &
 -  \frac{g^4}{2N_c} \, \delta^{ab} \int\limits_{q_1}\int\limits_{q_2}
\frac{e^{i{\vec b}_\perp\cdot ({\vec q_1}+{\vec q_2})}}{q_1^2 q_2^2} 
\left[ e^{i \frac{\vec r}{2}\cdot ({\vec{q_1}}-{\vec{q_2}})}
-\frac{1}{2} e^{i(q_1+q_2)\cdot \frac{\vec r}{2}}
  -\frac{1}{2} e^{-i(q_1+q_2)\cdot \frac{\vec r}{2}}\right]\nonumber\\
  &\times&  \la \trho^a(\vec q_1) \trho^b(\vec q_2) \ra_{\vec K_\perp} ~.
\eea

Multiplying both l.h.s and r.h.s by $2N_c$ to obtain  ${\cal
  T}^{O(\rho^2)}$, we can then perform the integration over impact
parameter in \eq{eq:Onium-amplitude} to obtain
\bea
\int d^2 b_\perp e^{i {\vec b}_\perp\cdot \vKT}~{\cal
  T}^{O(\rho^2)}(\vec r,\vec b_\perp; \vec K_\perp) &=& 2\,N_c\,\left[
 -\frac{g^4}{2 N_c}\,\int_{q_1} \frac{1}{q_1^2 (\vec q_1+ \vec
   K_\perp)^2}\,\left( 
e^{i \frac{{\vec r}}{2}\cdot (2 {\vec q_1}+\vKT)} 
- \cos\left(\frac{{\vec r}\cdot \vKT}{2}\right)\right) \right.\nonumber \\
&\times& \la \trho^a ({\vec q_1})\trho^a(-{\vec
  q_1}-\vKT)\ra_{K_\perp}
\Bigg] ~.
\eea
Defining the l.h.s of the above expression to be the Pomeron amplitude
${\cal P}(\vec r,\vec K_\perp)$ and replacing $\langle \trho(\vec
q_1)\trho(-{\vec 
  q_1}-\vKT)\rangle_{\vec K_T}$ on the r.h.s by the Pomeron form factor in
\eq{eq:Pom-FF1}, we obtain\footnote{In the forward scattering $\vec
  K_T\to 0$ limit, replacing $\frac{1}{2}\cG(\vec q,0)\rightarrow
  \bar{\mu}_{\rm MV}^2 \Theta(q^2-\Lambda^2)$, as discussed
  previously, reproduces the MV model expression
\bea \la 1- U(\vec x_T) U^\dagger(\vec y_T)\ra_{K_\perp=0} &=&
\frac{g^4C_F}{2} \int\limits_q \frac{1}{q^4} \left[e^{i\vec q\cdot \vec r }
  -1\right] \bar{\mu}^2_{\rm MV} \, \Theta(q^2-\Lambda^2) + \cdots
\eea . }, 
\bea
{\cal P}(\vec r,\vec K_\perp) = 2\,N_c \left[
-\frac{g^4 C_F}{2}\,\int_{q_1} \frac{1}{q_1^2 \, (\vec q_1+ \vec
  K_\perp)^2}\,\left(
e^{i \frac{{\vec r}}{2}\cdot (2 {\vec q_1}+\vKT)} 
- \cos\left(\frac{{\vec r}\cdot \vKT}{2}\right)\right) 
\cG({\vec q_1},-{\vec q_1}-\vKT)\right]~.   \label{eq:Mijtilde_Og2}
\eea
Here $C_F= (N_c^2-1)/2N_c$ is the
quadratic Casimir in the fundamental representation.

\begin{figure}
  \centering
   \includegraphics[width=0.45\textwidth]{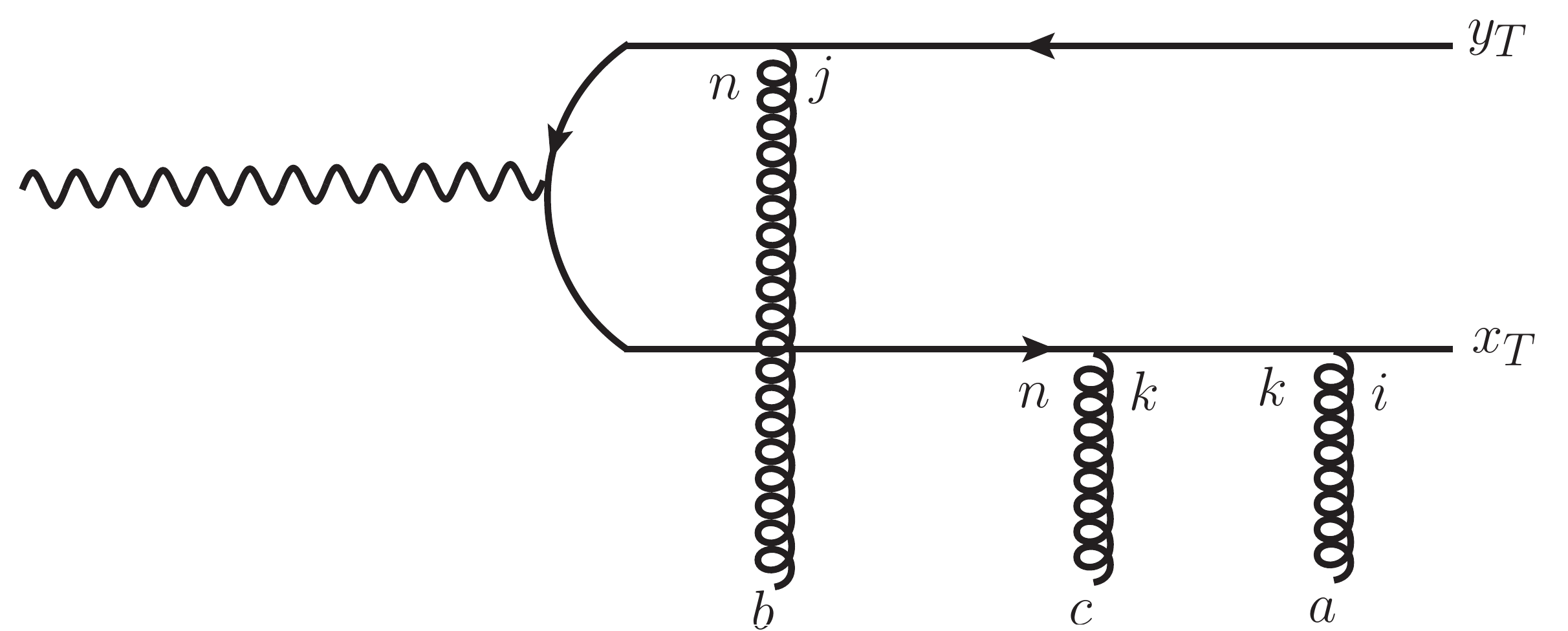}
  \caption{Illustration of the three-gluon contribution to the DIS exclusive amplitude for $c{\bar c}$ production.}
\label{fig:three-gluon}
\end{figure}

The amplitude for exclusive quarkonium production in DIS can also receive a contribution from three-gluon exchange, as illustrated in Fig.~\ref{fig:three-gluon}. 
This contribution is recovered in our approach by expanding \eq{eq:Mij(ell,P)} to $O(\rho^3)$. We begin by formally rewriting 
\bea
{\cal T} (\vec r,\vec b_\perp; \vec K_\perp) &=& 2\, N_c
\Bigg[ 1 - \frac{1}{2 N_c} {\rm tr} \left(\la U\left(\vec b_\perp +
  \frac{\vec r}{2}\right) U^\dagger\left( \vec b_\perp - \frac{\vec r}{2}\right)
  \ra_{\vec K_\perp} +
  \la U \left(\vec b_\perp - \frac{\vec r}{2}\right)
  U^\dagger\left( \vec b_\perp + \frac{\vec r}{2}\right)\ra_{\vec
    K_\perp}\right) \nonumber\\
&-& \frac{1}{2 N_c} {\rm tr} \left(\la U\left(\vec b_\perp +
  \frac{\vec r}{2}
  \right) U^\dagger\left( \vec b_\perp - \frac{\vec r}{2}\right)
 \ra_{\vec K_\perp} - \la U \left(\vec b_\perp - \frac{\vec
   r}{2}\right)
 U^\dagger\left( \vec b_\perp + \frac{\vec r}{2}\right)\ra_{\vec
   K_\perp}\right) \Bigg]\, .
 \eea
as the sum of a piece that's symmetric under $\vec b_\perp +
\frac{\vec r}{2}\leftrightarrow \vec b_\perp - \frac{\vec r}{2}$ and a piece
that is antisymmetric under this exchange. Expanding out both the
symmetric and antisymmetric terms to $O((gA^+)^3)$, or equivalently
$O(\rho^3)$, we find that the symmetric piece is identically zero at
this order. In other words, its impossible to have color-singlet
three-gluon exchange that is even under parity. The contribution of
the surviving term can be expressed as the Odderon amplitude
\bea
 i {\cal O}(\vec r,\vKT) &=& \int d^2 b_\perp e^{i {\vec b}_\perp\cdot
   \vKT}~{\cal T}^{O(\rho^3)}(\vec r,\vec b_\perp; \vec K_\perp)\,,
\label{eq:Odderon-amplitude}
\eea
where
\be
   {\cal T}^{O(\rho^3)}(\vec r, \vec b_\perp; \vec K_\perp) =
   \frac{1}{2 N_c} {\rm tr} \left(\la U\left(\vec b_\perp + \frac{\vec
     r}{2}\right)
   U^\dagger\left( \vec b_\perp - \frac{\vec r}{2}\right)
 \ra_{K_\perp} - \la U \left(\vec b_\perp - \frac{\vec
   r}{2}\right)
 U^\dagger\left( \vec b_\perp + \frac{\vec r}{2}\right)\ra_{\vec K_\perp}\right)
\ee
has the form of the expectation value of the Odderon
operator~\cite{Hatta:2005as}.

Working the r.h.s out to cubic order in $gA^+$ (or equivalently
$\rho$)--see Appendix B for details--one obtains
\bea
    {\cal T}^{O(\rho^3)}({\vec r}, {\vec b_\perp}; \vec K_\perp) &=&
    -\frac{g^6}{8N_c} d^{abc}
\int\limits_{q_1}\int\limits_{q_2}\int\limits_{q_3}
\frac{1}{q_1^2}\frac{1}{q_2^2}\frac{1}{q_3^2}\,
\la\trho^a(\vec q_1) \trho^b(\vec q_2)\trho^c(\vec q_3)\ra_{K_\perp}\,
e^{i {\vec b_\perp}\cdot ({\vec q_1}+{\vec q_2}+{\vec q_3})}   \nonumber\\
&\times & 
\left[ 2 \sin\left(\frac{{\vec r}}{2}\cdot ({\vec q_1}-{\vec q_2}-{\vec q_3})\right) + \frac{2}{3}\sin\left(\frac{{\vec r}}{2}\cdot ({\vec q_1}+{\vec q_2}+{\vec q_3})\right)\right] ~.
\eea
Note that only the terms proportional to $\sim d^{abc}$
from $\la\rho^a(\vec q_1) \rho^b(\vec q_2)\rho^c(\vec q_3)\ra_{K_\perp}$
contribute. Further, employing our definition of the Odderon amplitude in \eq{eq:Odderon-form-factor}, and using the identity
\be
d^{abc} d^{abe} = \frac{N_c^2-4}{N_c}\, \delta^{ce}\,,
\ee
we obtain the Odderon amplitude to be 
\bea
i{\cal O}(\vec r;\vec K_\perp) &=& -g^6\frac{(N_c^2-4)(N_c^2-1)}{8N_c^3}
\int\limits_{q_1}\int\limits_{q_2}
\frac{1}{q_1^2}\frac{1}{q_2^2}\frac{1}{(\vec q_1+\vec q_2+\vec
  K_\perp)^2}\,
\cG_O(\vec q_1,\vec q_2,-\vec K_\perp-\vec q_1-\vec q_2;\vec K_\perp) \nonumber\\
& \times &  \left[ 2 \sin\left(\frac{{\vec r}}{2}\cdot (2{\vec q_1}+\vKT)\right) - \frac{2}{3}\sin\left(\frac{{\vec r}}{2}\cdot \vKT\right)\right] ~\,,
\label{eq:Odderon-operator}
\eea
where $\cG_O(\vec q_1,\vec q_2,-\vec K_\perp-\vec q_1-\vec q_2; \vec
K_\perp)$ is the Odderon form factor from
\eq{eq:Odderon-form-factor} and $(N_c^2-1)(N_c^2-4)/4\,N_c^2 = C_{3F}$
is the cubic Casimir constant of SU($N_c$) in the fundamental
representation.

The Odderon expectation value $iO(\vec x_T,\vec y_T; K_T=0)$ in the forward
limit has been computed previously in the MV model, where the weight
functional (appropriately normalized) describing the distribution of
color charges in a large nucleus has the general
form~\cite{Jeon:2004rk,Jeon:2005cf}\footnote{A quartic term $\sim
  \rho^a(\vec x_\perp) \rho^a(\vec x_\perp) \rho^b(\vec x_\perp)
  \rho^b(\vec x_\perp)/\kappa_4$ arises too~\cite{Petreska_rho4}; it
  ensures that the action for $\rho$ is bounded from below.},
\be
W[\rho] = \int [d\rho]\,\exp\left( -\int d^2 x_\perp
\left[\frac{\rho^a(\vec x_\perp)\rho^a(\vec x_\perp)}{2\mu^2} -
  \frac{d_{abc}\, \rho^a(\vec x_\perp) \rho^b(\vec x_\perp)
    \rho^c(\vec x_\perp)}{\kappa_A}\right]\right)\,.
\ee
The cubic Casimir term here has the weight $\kappa_A =g^3 A^2
N_c/\pi^2 R^4$ and will of course give a non-zero value for the Odderon
form factor. For a large nucleus, if the typical magnitude of
$\rho\sim \sqrt{\mu_{\rm MV}^2}\sim A^{1/6}$, this cubic Odderon term
is subleading relative to the quadratic Pomeron term in $W[\rho]$ by
$A^{-1/6}$, which is a weak suppression factor even for a large
nucleus. The expectation value $iO(x_T,y_T)$ of the Odderon operator
computed in the MV model gives
\be
iO(\vec x_T,\vec y_T) = \alpha_S^3 \frac{(N_c^2-4)(N_c^2-1)}{4\pi r_0^2 N_c^3}\,
A^{1/3}\,\int d^2{\bf u}\, \ln^3\frac{|{\bf x}-{\bf u}|}{|{\bf y}-{\bf
    u}|}\, ,
\label{eq:MV-odderon}
\ee
where $r_0=1.12$ fm. This expression is also recovered in a
perturbative QCD computation~\cite{Kovchegov:2003dm}.  We can compare this expression to
\eq{eq:Odderon-operator}, for $A\rightarrow 1$ and in the forward
limit of $\vec K_T\rightarrow 0$. As discussed in \cite{Jeon:2005cf}, the logarithm above can be expressed in terms of the 
Coulomb propagator in two dimensions. Making use of this fact, we observe that \eq{eq:MV-odderon}  can be reexpressed as \eq{eq:Odderon-operator} 
if the Odderon form factor  $\cG_O$ is a constant everywhere except in the infrared due to the previously discussed constraint from color neutrality. 
Conversely, the structure of $\cG_O$ in \eq{eq:Odderon-operator}, and hence the Odderon operator at large $x_{\rm Bj}$, can be very different from the expectation from the MV model.

\subsection {Cross-section for exclusive production of $J/\Psi$ and $\eta_c$ mesons at large $x_{\rm Bj}$}

The general formalism for exclusive quarkonium production that we
outlined in the previous section can now be adapted to compute the
cross-section for specific quarkonium states.  We will consider here
the $J/\Psi$ because it is the most easily accessible Onium state, and
the $\eta_c$ because it is the lightest state with unique features
that promise novel insight into nonperturbative QCD. Since we are
interested in many-body color charge correlators of valence Fock
states in this work, our discussion is most relevant for exclusive
production of these quarkonium states at large $x_{\rm Bj}$. As noted,
this is a regime that is already accessible with the high luminosity
DIS experiments at Jefferson Lab and at   a future   EIC.\\

The cross-section for exclusive $J/\Psi$ production can be expressed as 
\be
\frac{d\sigma_{T,L}^{\gamma^*p\rightarrow J/\Psi \, p}}{dt} = \frac{1}{16\pi} |\,{\cal A}_{T,L}^{\gamma^*p\rightarrow J/\Psi p}\,|^2 \, ,
\ee
where 
\be \label{eq:ATL_JPsi_P_iO}
{\cal A}_{T,L}^{\gamma^*p\rightarrow J/\Psi p} (Q^2,\vKT)\sim i \int
\dd^2r \int \frac{dz}{4\pi}
\, \left(\Psi_{\gamma^*}
\Psi^*_{J/\Psi}\right)(\vec r,z,Q^2)~e^{-i\frac{(1-2z)}{2}\vec r\cdot
  \vKT} \,
\left[{\cal
    P}(\vec r,\vec K_\perp) + i{\cal O}(\vec r,\vec K_\perp)\right]\,.
\ee
Here $K_\perp^2 = -t$, and $\Psi_{J/\Psi}$, $\Psi_{\gamma^*}$ denote
the $J/\psi$ and virtual photon light cone wave functions (for
longitudinal or transverse polarization); their product is summed over
the helicities of the $c$ and $\bar c$ quarks.  Further, ${\cal P}$ is the
Pomeron contribution to the exclusive $J/\Psi$ amplitude given in
\eq{eq:Mijtilde_Og2} and $i{\cal O}$ is the respective Odderon
contribution given by \eq{eq:Odderon-operator}. The former is directly
proportional to the Pomeron color charge form factor and the
latter to the Odderon color charge form factor. These two terms in
${\cal A}_{T,L}^{\gamma^* p\rightarrow J/\Psi \,p}$ contain the
important QCD physics underlying the Regge theory based descriptions
of elastic/exclusive cross-sections in terms of imaginary and real
terms respectively~\cite{Forshaw:1997dc}. There is an additional
kinematic contribution coming from the non-zero values of $\Delta x$
discussed in Section II; however, as we demonstrate in Appendix A,
these contributions are $1/P^+$ suppressed.

Some remarks on the contribution due to the Odderon are in
order. $i{\cal O}$ is odd under charge conjugation, which corresponds
to the simultaneous transformations $\vec{r}\to -\vec{r}$, $z\to
1-z$. On the other hand, $\Psi_{\gamma^*} \Psi^*_{J/\Psi}$ has even C
parity. Therefore, the integral over $i{\cal O}$ in
\eq{eq:ATL_JPsi_P_iO} is non-zero only if the final state is
restricted to, for example, $p^+_c < p^+_{\bar c}$ ($z<1/2$). This
prevents the cancellation of the amplitude with its C
conjugate. Likewise, the Odderon contribution to the above amplitude
will not cancel against its parity transform if the direction of the
momentum transfer $\vec K_\perp$ is fixed. The role of such charge
asymmetry and kinematic constraints in Pomeron-Odderon DIS amplitudes
has been noted previously for other final
states~\cite{Brodsky:1999mz,Hagler:2002nf}.

The two-gluon Pomeron and three-gluon Odderon form factors were
discussed previously in \cite{Brodsky:2000zc} albeit this work did not
identify these form factors as such. More importantly, we have
provided explicit first-principles expressions for the Pomeron form
factor in \eq{cgdef} and likewise for the Odderon form factor in
\eq{eq:Odderon-form-factor} in terms of the QCD light front
wavefunction for valence Fock states. Therefore exclusive measurements
of the $J/\Psi$ at large $x_{\rm Bj}$ offer the opportunity to extract
fundamental nonperturbative QCD physics contained in these
wavefunctions.

It is important to note that by large $x_{\rm Bj}$, we have $x_{\rm
  Bj}\approx 0.1$ in mind. At larger values of $x_{\rm Bj}$, our
approximations ignoring $\Delta x$ are no longer tenable. At smaller
values of $x_{\rm Bj} < 0.1$, higher gluon Fock states become
important. While these can be incorporated in our approach, and
matched eventually to the CGC EFT framework, their treatment is
outside the scope of the present discussion. \\

We observed that that while the exclusive $J/\Psi$ cross-section is
dominated by the Pomeron contribution, it can in principle be
sensitive to the Odderon form factor for particular kinematics. In
contrast to the $J/\psi$ however, the $\eta_c$ meson with its $P=-1$
and $C=+1$ quantum numbers, is dominantly produced in exclusive DIS by
the three-gluon color singlet Odderon exchange contribution. The
exclusive $\eta_c$ production amplitude is simply
\be
{\cal A}_{T,L}^{\gamma^*p\rightarrow \eta_c p} (Q^2,\vKT)\sim i \int
\dd^2r \int \frac{dz}{4\pi}
\, \left(\Psi_{\gamma^*} \, \Psi^*_{\eta_c}\right)(\vec r,z,Q^2)~
e^{-i\frac{(1-2z)}{2}\vec r\cdot \vKT}  \, i{\cal O}(\vec r,\vec K_\perp)\,,
\ee
where $\Psi_{\eta_c}$ is the light cone $\eta_c$ wavefunction. Indeed,
exclusive $\eta_c$ was proposed some time ago~\cite{Schaefer} as the
cleanest channel for discovery of the Odderon\footnote{For a nice
  review of both the theoretical work on the Odderon and experimental
  searches, we refer the reader to \cite{Ewerz:2003xi}.} where the
focus there was on $\eta_c$ production at small $x_{\rm Bj}$ at
HERA. Ref.~\cite{Engel:1997cga} followed the approach of
\cite{FukugitaKwiecinski,Czyzewski:1996bv} to estimate the HERA DIS
$\eta_c$ cross-section to be 47~pb for photo-production and 11~pb for
$Q^2=5$ GeV$^2$. However \cite{FukugitaKwiecinski,Czyzewski:1996bv}
express the Odderon form factor in terms of that of the Pomeron form
factor.  Our study shows that this assumption is likely unjustified;
we plan to investigate its quantitative impact in a future
publication. \\

Searches at HERA did not reveal any evidence for exclusive
$\eta_c$. From the theory perspective, this may be because the Odderon
amplitude is suppressed at small $x$. While not definitive,
studies of the small $x$ evolution of the Odderon suggest that its
energy dependence is much smaller than that of the
Pomeron~\cite{Bartels:1999yt}; it may even decrease with increasing
energy~\cite{Janik:1998xj,Lappi:2016gqe}. Therefore, searches at
larger values of $x_{\rm Bj}$ may be more promising. Further, since
the cross-section for such exclusive processes is small, such searches will
benefit from the much higher luminosities at Jefferson Lab and in
future at the EIC.

\section{Summary and Outlook}

In this paper, we developed a novel formalism within the framework of
light front QCD to compute color charge correlators and their
associated color charge form factors. For simplicity, we constructed
the quadratic $\langle\rho\rho\rangle$ and cubic $\langle
\rho\rho\rho\rangle$ correlators of valence quark Fock states in the
proton. The extension of our computation to include gluon and sea
quark color charge densities is straightforward if more
involved. These quadratic and cubic color charge correlators are
precisely the color singlet two-gluon Pomeron form factor and the
three-gluon Odderon form factor respectively. They capture important
nonperturbative physics on the spatio-temporal distribution of color
charges in the proton, and offer a complementary description of this
tomography to that offered by TMDs and GPDs. Further, they provide
useful classical intuition at the level of the Yang-Mills dynamics of
QCD. As a striking example, note that the Wong equations~\cite{Wong:1970fu} satisfied by
classical color charges in background gauge fields are embedded in the
structure of the QCD effective action~\cite{JalilianMarian:1999xt}. Classical intuition at this
level can motivate experimental searches for novel QCD effects.\\

While expressing observables in terms of expectation values of color charge correlators is
uncommon at large $x$ (see however \cite{Burkardt:2003yg}), it is a key
feature of the Color Glass Condensate framework at small $x$, whereby
dynamical many-body information from nonperturbative initial
conditions is encoded in a gauge invariant density matrix
$W[\rho]$. For a large nucleus, this quantity is the Gaussian weight
functional of the McLerran-Venugopalan model. However, this formalism
breaks down for the proton at large $x$ and the initial conditions for
small $x$ evolution of color charge correlations in the proton have a
significant source of uncertainty arising from the initial conditions.\\

We showed that exclusive measurements of quarkonia at large $x$ allow
for independent extraction of $\langle \rho\rho\rangle$ and
$\langle\rho\rho\rho\rangle$. Expectation values of these, and the
associated Pomeron and Odderon color charge form factors, can be
extracted from clean exclusive DIS measurements of quarkonium final
states at large $x$. These form factors, and in principle higher
moments of the color charge density, therefore provide a bridge
between small $x$ and large $x$ in QCD, one that is constrained by
high energy proton-proton and proton-nucleus experiments on
multiparticle correlations at RHIC and and LHC on the one hand, and
DIS experiments at Jefferson Lab on the other.   We also applied the formalism towards computing the gluon distribution  of a proton, and obtained sensible results.
We anticipate that the
Electron-Ion Collider, which will have an unparalleled combination of
$x$ reach and high luminosities, will bring powerful new insight into
the underlying dynamics of many-body color charge correlations in QCD.\\

Another interesting avenue of research that presents itself is the extraction of color charge correlations and form factors in polarized deep inelastic scattering and polarized proton-proton collisions. Odderon exchange  can for instance be probed in the single spin asymmetries measured in polarized proton-proton collisions~\cite{Kovchegov:2012ga}. Single spin asymmetries in semi-inclusive open charge production in polarized DIS are also sensitive to the Odderon operator~\cite{Zhou:2013gsa,Dong:2018wsp}. These connections between color charge form factors in a wide range of experiments are ripe for further exploration.

\section*{Acknowledgements}

This material is based on work supported by the U.S. Department of
Energy, Office of Science, Office of Nuclear Physics, under Contracts
No. DE-SC0012704 (A.D and R.V) and within the framework of the TMD
Theory Topical Collaboration (R.V). A.D.\ also acknowledges support by
the DOE Office of Nuclear Physics through Grant
No.\ DE-FG02-09ER41620; and from The City University of New York
through the PSC-CUNY Research grant 60262-0048.  A.D.\ would also like
to thank the Nuclear Theory Group of Brookhaven National Laboratory
for kind hospitality. G.A.~Miller would like to thank the Lab for
Nuclear Science at MIT, the Southgate Fellowship of Adelaide
University (Australia), the Bathsheba de Rothchild Fellowship of
Hebrew University (Jerusalem), the Shaoul Fellowship of Tel Aviv
University, the Physics Division of Argonne National Laboratory and
the U.\ S.\ Department of Energy Office of Science, Office of Nuclear
Physics under Award Number DE-FG02-97ER-41014 for support that enabled
this work.  This work was also supported by the U.\ S.\ Department of
Energy Office of Science, Office of Nuclear Physics under Award
Numbers DE-FG02-97ER-41014, DE-FG02-94ER40818 and DE-FG02-96ER-40960,
the Pazy foundation, and by the Israel Science Foundation (Israel)
under Grants Nos. 136/12 and 1334/16.\\

We thank D.~Kharzeev, L.~Motyka  and T.~Stebel for useful comments on the manuscript.
The figures in this paper have been prepared with Jaxodraw~\cite{jaxo}.

\section*{Appendix A: $P^+$ limit of the density operator, and the
  parton pancake}
Consider \eq{densop}. This contains a term:
\bea P^+ e^{i(x_q-x_p)P^+ r^-},\label{term}\eea
which oscillates like crazy if $P^+\to\infty$ unless $r^-$  or/and $(x_p-x_q) $ vanishes. In those cases the term is infinite. This is suggestive of delta functions, and the pancake  shape of high energy projectiles. 

To better understand the term \eq{term} consider a test function $f(r^-)$ which is continuous at the  origin and non-zero over a finite region of space. Such would arise in taking the matrix element of the density operator in the proton wave function.
Then 
\bea&\int dr^- f(r^-)  \lim_{P^+\to\infty} P^+ e^{i(x_q-x_p)P^+ r^-}\nonumber\\&
=  \lim_{P^+\to\infty}  \int du e^{i(x_q-x_p)u}f(u/P^+)\nonumber\\&=f(0)\int du e^{i(x_q-x_p)u} =f(0)2\pi \delta(x_q-x_p)\label{dd}
.\eea
Thus the term of \eq{term} and the density operator of \eq{densop}  acts as a delta function in both $x_q-x_p$ and $r^-$. \\

Thus effectively
\bea \lim_{P^+\to\infty} P^+ e^{i(x_q-x_p)P^+ r^-}\rightarrow 2\pi \delta(x_p-x_q)\delta(r^-).\label{delta}\eea
We therefore see that the density $\rho^a(r)$ contains $ \delta(r^-)$, hence the pancake shape.  Using \eq{delta} in \eq{densop}  and integrating over $r^-$ leads immediately to \eq{rhor}. \\

The corrections of order $1/P^+$ can be understood from \eq{dd}, by using
\bea f(u/P^+)\approx f(0)+ f'(0){u\over P^+}.\eea
Including the second term gives a correction term:
\bea
{2\pi i\over P^+} f'(0){\partial\over \partial x_q}\delta(x_q-x_p)
.\eea

\section*{Appendix B: The Odderon amplitude in terms of the Odderon form factor}
The Odderon contribution to the amplitude in \eq{eq:Odderon-amplitude} can be written our explicitly as 
\bea
&& \frac{1}{2N_c}\,\text{tr}~\la
  U(\vec x_T) U^\dagger(\vec y_T) -
  U(\vec y_T) U^\dagger(\vec x_T)\ra_{K_\perp}=\\
&&\frac{ig^3}{2N_c}\,\text{tr}
 \left<\int dx^- \int dy^- \int^{y^-} dz^- A^+(x^-,\vec x_T)
   A^+(y^-,\vec y_T) A^+(z^-,\vec y_T) \right.\label{eq:O_line1}\\
  &-& \int dx^- \int dy^- \int^{y^-} dz^- A^+(x^-,\vec y_T)
A^+(y^-,\vec x_T) A^+(z^-,\vec x_T) \label{eq:O_line2}\\
&-&  \int dx^- \int_{x^-} dy^- \int dz^- A^+(x^-,\vec x_T)
A^+(y^-,\vec x_T) A^+(z^-,\vec y_T) \label{eq:O_line3}\\
&+& \int dx^- \int_{x^-} dy^- \int dz^- A^+(x^-,\vec y_T)
A^+(y^-,\vec y_T) A^+(z^-,\vec x_T) \label{eq:O_line4}\\
&+& \int dx^- \int_{x^-} dy^- \int_{y^-} dz^- A^+(x^-,\vec x_T)
A^+(y^-,\vec x_T) A^+(z^-,\vec x_T) \label{eq:O_line5}\\
&-&  \int dx^- \int_{x^-} dy^- \int_{y^-} dz^- A^+(x^-,\vec y_T)
A^+(y^-,\vec y_T) A^+(z^-,\vec y_T) \label{eq:O_line6}\\
&-& \int dx^- \int^{x^-} dy^- \int^{y^-} dz^- A^+(x^-,\vec y_T)
A^+(y^-,\vec y_T) A^+(z^-,\vec y_T) \label{eq:O_line7}\\
&+&\left. \int dx^- \int^{x^-} dy^- \int^{y^-} dz^- A^+(x^-,\vec x_T) A^+(y^-,\vec x_T) A^+(z^-,\vec x_T) \right>_{K_\perp}~. \label{eq:O_line8}
\eea
With a little algebra one can combine \eq{eq:O_line1}
and \eq{eq:O_line4} to
\bea
\text{tr}~t^a t^b t^c & &
\int dx^-\int dy^-\int dz^-
A^{+a}(z^-,x_T)A^{+b}(x^-,y_T)A^{+c}(y^-,y_T) \nonumber\\
&=& \frac{1}{4}d^{abc} \int dx^-\int dy^-\int dz^- A^{+a}(z^-,x_T)
A^{+b}(x^-,y_T) A^{+c}(y^-,y_T) ~.
\eea
In the last step we have used that the factor multiplying $\text{tr}~
t^a t^b t^c$ is symmetric under $b\leftrightarrow c$.  Since all
fields are now integrated over $x^-$ without limits they can be traded
for $\rho^a(q)$ from eq.~(\ref{eq:rho^a_k}) so that the previous line
becomes
\bea
\frac{g^3}{4}d^{abc}
\int\limits_{q_1}\int\limits_{q_2}\int\limits_{q_3}
\frac{1}{q_1^2}\frac{1}{q_2^2}\frac{1}{q_3^2}\,
e^{i(q_1\cdot x_T+(q_2+q_3)\cdot y_T)}\,
\rho^a(q_1)\rho^b(q_2)\rho^c(q_3)~.
\eea
Along the same lines, the sum of (\ref{eq:O_line2}) and (\ref{eq:O_line3}) can be rewritten as
\bea
& & -\frac{1}{4}d^{abc}
\int dx^-\int dy^-\int dz^- A^{+a}(z^-,y_T)A^{+b}(x^-,x_T)A^{+c}(y^-,x_T) \nonumber\\
&=& 
-\frac{g^3}{4}d^{abc}
\int\limits_{q_1}\int\limits_{q_2}\int\limits_{q_3}
\frac{1}{q_1^2}\frac{1}{q_2^2}\frac{1}{q_3^2}\,
e^{i(q_1\cdot y_T+(q_2+q_3)\cdot x_T)}\,
\rho^a(q_1)\rho^b(q_2)\rho^c(q_3)~.
\eea
The remaining terms from eqs.~(\ref{eq:O_line5}-\ref{eq:O_line8}) involve integrals over $A^+(x^-,x_T)$ at the same point $x_T$, i.e.\ integrals of the {\em same} (matrix valued) function $A^+(x^-)$. One may thus use standard identities for ``time'' ordered exponentials of a matrix $A(t)$:
\be
T~\int dt_1 \cdots \int dt_n A(t_1) \cdots A(t_n) = T~\int dt_1 \cdots \int dt_n \frac{1}{n!} \, \sum_{\text perm} A(t_{i_1}) \cdots A(t_{i_n})~.
\ee
The sum is over all permutations of $A(t_1)$, $A(t_2), \cdots, A(t_n)$. We
can now express~(\ref{eq:O_line5})+(\ref{eq:O_line8}) as
\bea
& & 2 \frac{1}{3!} \frac{1}{4} d^{abc}
\int dx^- \int dy^- \int dz^- A^{+a}(x^-,\vec x_T)
A^{+b}(y^-,\vec x_T) A^{+c}(z^-,\vec x_T) \nonumber\\
&=& \frac{g^3}{12} d^{abc} \int\limits_{q_1}\int\limits_{q_2}\int\limits_{q_3}
\frac{1}{q_1^2}\frac{1}{q_2^2}\frac{1}{q_3^2}\,
e^{i(q_1+q_2+q_3)\cdot x_T}\,
\rho^a(q_1)\rho^b(q_2)\rho^c(q_3)~.
\eea
Similarly, the sum of (\ref{eq:O_line6}) and (\ref{eq:O_line7}) is given by
\bea
& & - \frac{1}{12} d^{abc}
\int dx^- \int dy^- \int dz^- A^{+a}(x^-,\vec y_T)
A^{+b}(y^-,\vec y_T) A^{+c}(z^-,\vec y_T) \nonumber\\
&=& -\frac{g^3}{12} d^{abc} \int\limits_{q_1}\int\limits_{q_2}\int\limits_{q_3}
\frac{1}{q_1^2}\frac{1}{q_2^2}\frac{1}{q_3^2}\,
e^{i(q_1+q_2+q_3)\cdot y_T}\,
\rho^a(q_1)\rho^b(q_2)\rho^c(q_3)~.
\eea
%

\end{document}